\documentclass[reprint,article,showpacs,preprintnumbers,amsmath,amssymb,citeautoscript,aip,superscriptaddress]{revtex4-1}
\usepackage{graphicx,wasysym}
\usepackage{dcolumn}
\usepackage{bm}
\usepackage{hyperref}
\usepackage{color}
\setcitestyle{super}

\usepackage{color}

\newcolumntype{P}[1]{>{\centering\arraybackslash}p{#1}}

\makeatletter

\newcommand{\thickhline}{%
    \noalign {\ifnum 0=`}\fi \hrule height 0.4pt
    \futurelet \reserved@a \@xhline}
\newcolumntype{"}{@{\hskip\tabcolsep\vrule width 1pt\hskip\tabcolsep}}
\makeatother

\newcommand{\avg}[1]{\left \langle #1 \right \rangle}

\begin{document}

\preprint{1}

\begin{abstract}
We present a means of studying rare reactive pathways in open quantum systems using Transition Path Theory and ensembles of quantum jump trajectories. This approach allows for elucidation of reactive paths for dissipative, nonadiabatic dynamics when the system is embedded in a Markovian environment. 
We detail the dominant pathways and rates of thermally activated processes, as well as the relaxation pathways and photoyields following vertical excitation in a minimal model of a conical intersection. We find that the geometry of the conical intersection affects the electronic character of the transition state, as defined through a generalization of a committor function for a thermal barrier crossing event. Similarly, the geometry changes the mechanism of relaxation following a vertical excitation. Relaxation in models resulting from small diabatic coupling proceed through pathways dominated by pure dephasing, while those with large diabatic coupling proceed through pathways limited by dissipation. The perspective introduced here for the nonadiabatic dynamics of open quantum systems generalizes classical notions of reactive paths to fundamentally quantum mechanical processes. 
\end{abstract}

\title{Nonadiabatic transition paths from quantum jump trajectories}

\author{Michelle C. Anderson}
\affiliation{%
Department of Chemistry, University of California, Berkeley, California 94720, USA \looseness=-1
}
\author{Addison J. Schile}
\affiliation{%
Department of Chemistry, University of California, Berkeley, California 94720, USA  \looseness=-1
}
\author{David T. Limmer} \email{dlimmer@berkeley.edu}
\affiliation{%
Department of Chemistry, University of California, Berkeley, California 94720, USA  \looseness=-1
}
\affiliation{%
Kavli Energy NanoSciences Institute, University of California, Berkeley, California 94720, USA \looseness=-1
}\affiliation{%
Chemical Sciences Division, Lawrence Berkeley National Laboratory, Berkeley, California 94720, USA\looseness=-1
}
\affiliation{%
Materials Sciences Division, Lawrence Berkeley National Laboratory, Berkeley, California 94720, USA\looseness=-1
}

\date{\today}


\maketitle

\section*{Introduction}

Chemical reactions are largely understood through the identification and enumeration of reactive paths that evolve on potential energy surfaces and over transition states. Such a classical description of reactivity is valid when motion of heavy particles proceeds adiabatically with respect to the electronic degrees of freedom.\cite{peters2017reaction,nitzan} However if the nuclear and electronic degrees of freedom become correlated and the resultant dynamics are nonadiabatic, classical notions of reactive paths and spatially localized transition states break down.\cite{hanggi1990reaction,miller1974quantum,leggett1984quantum,voth_chandler}  Finding an appropriate framework for understanding reaction mechanisms when the dynamics are quantum mechanical remains an outstanding challenge.\cite{tully_2012} Here we provide such a framework in the limit that the quantum dynamics are mediated by a Markovian bath that acts to dephase superpositions and dissipate energy. To accomplish this, we generalize Transition Path Theory\cite{weinan_ve} developed in the context of classical Markov processes to study a system evolving through its Hilbert space with transitions mediated by a thermal bath. This construction provides a means of defining reactive paths, calculating rates, and determining yields of photochemical processes when the coupling between the system and the bath is weak and the dynamics are nonadiabatic. 

Nonadiabatic dynamics occur in photochemical reactions when excitation supplies reactants sufficient energy to access higher lying electronic states.\cite{schuurman_stolow} Molecular configurations where multiple adiabatic potential surfaces meet, such as conical intersections or conical seams,\cite{zhu2016non} necessarily result in a breakdown of the adiabatic or Born-Oppenheimer approximation yielding a dynamics that convolve nuclear and electronic motion.\cite{worth_cederbaum,levine2007isomerization} Conical intersections especially play important roles in natural and synthetic systems as they mediate ultrafast nonradiative relaxation.\cite{neville,matsika,curchod2018ab} In nucleotides they dissipate dangerous excitations\cite{barbatti_aquino} and in photoswitches they funnel excitation energy into directed conformational changes \cite{schultz2003mechanism}. Rates and yields of nonadiabatic reactions are sensitive to changes in the environment.\cite{schuurman_stolow,courtney1985photoisomerization,hicks1987polarity}
For example, different solvents can strongly influence the photoyields and mechanisms of azobenzene isomerization as inferred from pump-probe experiments.\cite{dhammika} When nonadiabatic dynamics occur in condensed phases, the dual importance of nonadiabatic and dissipative effects renders the description of their reactive dynamics necessarily a statistical one.\cite{schile2019simulating,halpern2020fundamental,chuang2021extreme} This requires a theoretical framework capable of describing the ensemble of reactive paths analogously to that developed within a classical context.

Understanding reaction dynamics in the nonadiabatic limit and in the presence of many interacting degrees of freedom is difficult, as many tools for analyzing reaction paths and transition states depend on classical views of the system or the Born-Oppenheimer approximation. Nevertheless, attempts to understand reactive dynamics of open quantum systems have resulted in the development of rigorous, quantum mechanical transition state theories based on time {\color{black}  correlation function formalisms.\cite{miller1974quantum,truhlar,hele_althorpe2013,hele_althorpe2013II} 
}
Such theories can account for low temperature, deep tunneling effects,\cite{wolynes,grote_hynes,gillan_1987,gillan_bc,voth_1993,voth_chandler,jang_voth,ranya_ananth} and rate expressions can be formulated within an instanton approximation to the real or imaginary time path integral\cite{voth_chandler,gillan_bc,hele_willatt2015}.
These theories, however, are typically focused on a single potential energy surface and not necessarily applicable to studying the complex behaviors around conical intersections. Notable exceptions are so-called nonadiabatic transition state rate expressions which use Fermi's golden rule with analytical continuation or nonadiabatic extensions to ring polymer instanton theory.{\color{black}\cite{ranya_ananth,heller2020instanton,lawrence2020general,hammes-schiffer,hammes-shiffer_tully,wolynes1987imaginary,litman2022dissipative,kretchmer2013direct,miller1983quantum,tao_shushkov2018,tao_shushkov2019,shushkov_li2012,lawrence_manolopoulos2019,richardson_thoss2013}
 Instanton and lowest order perturbation theories admit the description of only one dominant transition pathway, } however at elevated temperatures or for relaxation following photoexcitation, a
broad ensemble of paths is expected to contribute. {\color{black} Some recent work on path-integral based corrections to quantum transition state theory allows for inclusion of multiple pathways} \cite{fang_thapa2019,fang_zarotiadis2020,lawrence_manolopoulos2019,lawrence_manolopoulos2020}.

In classical systems, trajectory based approaches have been successful at distilling reactive dynamics in condensed phase systems through the introduction of transition path ensembles, and associated ensembles of transition states.\cite{peters2016reaction,geissler1999kinetic} Analogous approaches can be adapted to treat quantum systems when the uncertainty is only in the initial condition.\cite{dodin2019state,dodin2022trajectory} 
{\color{black}  
More generally, uncertainty results from both initial conditions as well as quantum superpositions, and developing an analogous trajectory ensemble based approach to address both is complicated by the inability to generally define a quantum trajectory under conditions demanding that real, positive probabilities are assigned to each trajectory in order to guarantee equivalent interpretation with a classical ensemble.\cite{wiseman_howard} }

A quantum trajectory conveying information of the state of the system at every time is analogous to repeated measurements of the system and as such will enforce an evolution which is incoherent.\cite{breuer_petruccione} In the limit of weak system bath coupling however, quantum jump trajectories can be defined as a time record of wavefunction evolution conditioned on a choice of measurement, with different methods of experimental observation corresponding to different stochastic unravellings of a reduced density matrix dynamics.\cite{garrahan_guta}  Quantum jumps have been witnessed and manipulated in the lab, with their counting statistics observed and even controlled.\cite{garrahan_guta,minev,carollo_jack,carollo_garrahan} Applications of trajectory based quantum control, such as randomly resetting the state of propagating trajectories or interfering to rotate away from dark states in {\color{black} Hilbert space, has gained interest, lending credence to quantum jump trajectories as a realizable process which provides insight into quantum phenomena\cite{minev,perfetto,garrahan_guta}.  However, multiple unravellings of the density matrix exist, and absent the direct invocation of explicit experimental design the use of any specific stochastic unravelling to aid in the interpretation of the behavior of a quantum system is subjective.  Nevertheless, extensions of quantum jumps to chemical dynamics problems have recently been developed\cite{schile_limmer} allowing the framework of transition path sampling\cite{chandler_dellago} to be brought to bare on questions of proton-coupled electron transfer and thermal barrier crossings, and utility derived within that perspective.}

Expanding on these previous approaches, here we generalize the Transition Path Theory (TPT)\cite{weinan_ve} of classical Markov models to nonadiabatic dynamics employing quantum jump trajectories as the generator of the underlying stochastic process. TPT supplies a number of formal results allowing evaluation of typical reaction paths, locations of dynamical bottlenecks and rate constants.\cite{metzner_schutte,noe_schutte} It has been used extensively in protein folding to reveal the structure of the folding landscape.\cite{noe_schutte,meng_shukla,yu_lapelosa,zheng_gallicchio}  {\color{black} Here, we use} TPT to characterize the behavior of dynamics in the vicinity of a conical intersection.  We use quantum jump dynamics to generate a Markov process between energy eigenstates in a linear vibronic coupling model\cite{kouppel1984multimode} and describe rare reactive dynamics.\cite{dellago1999calculation} { \color{black} The experimental equivalent to the jump unravelling employed would involve the measurement of phonons entering and exiting the surrounding bath, or the continuous monitoring of the system's energy.\cite{nguyen2014studying} } From probabilities to react and equilibrium populations of the system, we calculate reactive fluxes between each eigenstate and from the resulting graph of reactive fluxes, calculate reaction rates, quantum yields and principally important reaction pathways. We consider thermally activated dynamics and vertical excitations for a variety of conical intersection geometries. We observe an increase of the thermal barrier crossing rate with increasing diabatic coupling strength and identify a change in mechanism from tunneling at low diabatic coupling to traversing around the conical intersection at high diabatic coupling. The destination of relaxing trajectories following vertical excitation at low diabatic coupling was determined by dephasing effects with the trajectory's fate sealed at high energies, whereas the destination of relaxing trajectories at high diabatic coupling was decided at energies comparable to the barrier, with a larger photoyield and a greater variety of relaxation pathways.

\section*{Nonadiabatic reactions in open quantum systems}

In order to apply the framework of TPT, we require a dynamics that is dissipative and Markovian. For a quantum system in contact with an infinite heat bath, we focus on the dissipative evolution of a reduced density matrix evaluated by integrating out the bath degrees of freedom.  
We consider models defined through separable Hamiltonians written as the sum of an operator acting only on the system, $H_s$, an operator acting only on the bath, $H_b$, and an interaction term, $V$, which couples the two. \cite{nitzan} The full Hamiltonian, $H$, is 
\begin{equation}
    H = H_s + H_b + V
\end{equation}
where the coupling term is taken as bilinear in the system and bath operators
\begin{equation}
    V =  Q_t F_t + Q_c F_c
\end{equation}
where, $Q_{t/c}$ acts within the system's Hilbert space and $F_{t/c}$ in the bath's.\cite{pollard_friesner}
In the model of a conical intersection we study, we represent two vibrational modes explicitly, the tuning mode $Q_t$ and the coupling mode $Q_c$. All other vibrational modes are incorporated in the harmonic bath to which the system is coupled through the explicit modes.

\subsection*{Quantum jump dynamics}
In the limit that the system and the bath are weakly coupled, the dynamics of the reduced density matrix is Markovian and is well described by a quantum master equation.\cite{alicki2006internal} The specific dynamics we consider are those that result from the Lindblad master equation. The Lindblad master equation derives from second order perturbation theory applied to the system-bath coupling operator followed by the Markovian and secular approximations.\cite{breuer_petruccione} The time evolution of a reduced density matrix spanning the system Hilbert's space, $\rho(t)$, is given by a linear operator $\mathcal{D}$,
\begin{eqnarray}
    \frac{\partial \rho(t)}{\partial t} &=& \mathcal{D}[\rho(t)] 
\end{eqnarray}
which is decomposable into two types of terms
\begin{eqnarray}
\label{eq:D}
\mathcal{D}[\rho(t)]&=& -\frac{i}{\hbar} \left[ H_s, \rho(t) \right] + \\ 
        &&\sum_{a,i,j} \Gamma^a_{ij} \left( L^a_{ij} \rho(t) (L^a_{ij})^\dagger - \frac{1}{2} \left\{ (L^a_{ij})^\dagger L^a_{ij}, \rho(t) \right\} \right) \nonumber
\end{eqnarray}
where the sum over $i,j$ is over all of the eigenstates of the system, $a$ sums over the independent baths to which the system is coupled, and $\hbar$ is Planck's constant. The first term is a coherent portion of the dynamics determined by $H_s$, which generates a time evolution provided the density matrix is not in an energy eigenstate. The second term is an incoherent hopping process that reflects the influence of the bath and leads to irreversible relaxation of the system. The dissipative dynamics are determined by $\Gamma^a_{ij}$, the jump rates between states $i$ and $j$ associated with a jump operator $L^a_{ij}$ with Hermitian conjugate $(L^a_{ij})^\dagger$. Within the energy eigenbasis, the Lindblad equations represent a set of linearly coupled equations for the diagonal elements of the density matrix in the energy eigenstate representation.

Each jump operator, $L^a_{ij}$, is constructed from a projection matrix onto energy eigenvector subspaces in the energy eigenbasis, $\{| \phi \rangle\}$, and each Lindblad operator is determined from the system-bath coupling such that \cite{jeske_cole,schile_limmer}
\begin{equation}
    L_{ij}^a = (Q_a)_{ij} | \phi_i \rangle \langle \phi_j |
\end{equation}
 for $i \neq j$, where $(Q_a)_{ij}=  \langle \phi_i | Q_a | \phi_j \rangle$. For $i = j$, a single operator,
\begin{equation}
L_{ii}^a= \sum_j (Q_a)_{jj} | \phi_j \rangle \langle \phi_j |
\end{equation}
results in pure dephasing. The Lindblad equation is equivalent to secular Redfield theory\cite{breuer_petruccione,jeske_cole} when the jump rates, $\Gamma_{ij}^a$, associated with each operator are obtained from equilibrium bath correlation functions of the form 
\begin{equation}
\Gamma_{ij}^a = \int_0^\infty e^{-i \omega_{ij}t} \langle F_a(t) F_a(0) \rangle \; dt
\end{equation}
where $\omega_{ij}=(E_i -E_j) / \hbar$ is determined by the energy eigenvalues of the isolated system and $\avg{\dots}$ is the averaging operation. Hopping rates evaluated in this way obey detailed balance and ensure the proper thermalization  within the system Hilbert space.\cite{levy2021response} 

The Lindblad master equation can be viewed as the average evolution associated with a stochastic Schrodinger equation. The unravelled Lindblad equations describe the progression of individual wavefunction trajectories under a Poisson jump process in which wavefunctions transition instantaneously between eigenstates.\cite{plenio1998quantum} The stochastic formulation allows investigation of reaction pathways and mechanisms of individual trajectories rather than mere inspection of the average behavior of a trajectory ensemble through density matrix evolution.\cite{schile_limmer} The Lindblad equation may be disassembled into a stochastic equation of motion describing the evolution of a single wavefunction, $ \Psi (t)$, as
\begin{eqnarray}
    d | \Psi (t) \rangle &&= - \frac{i}{\hbar} \tilde{H}_s | \Psi (t) \rangle dt +  \\ &&\sum_{a,i,j} \left(  \frac{\sqrt {\Gamma_{ij}^a} L^a_{ij}}{ \langle \Psi(t) | \Gamma^a_{ij} (L^a_{ij})^\dagger L^a_{ij} | \Psi(t) \rangle} -1 \right) | \Psi (t) \rangle dN^a_{ij} \nonumber
\end{eqnarray}
where $dN_{ij}^a = 0,1$ and $(dN^a_{ij})^2 = dN^a_{ij}$ characterize the Poisson noise associated with the stochastic process. 
The effective Hamiltonian of the unravelled formulation
\begin{equation}
    \tilde{H}_s = H_s - \frac{i}{2} \sum_{i,j} \Gamma^a_{ij} (L^a_{ij})^\dagger L^a_{ij}
\end{equation}
includes an additional, anti-hermitian summation due to the effects of the jump operators. 
{\color{black} 
While the deterministic evolution does not preserve norm, the stochastic evolution restores it on average.
}
The evolution of the reduced density matrix, $\rho(t)= {\avg{| \Psi (t)\rangle \langle  \Psi(t)| }}$, can be recovered by averaging over a sufficient number of Lindblad trajectories. The unravelling of the Lindblad equation clarifies that within the eigenstate representation of the system, the evolution of the system is analogous to a classical, continuous time Markov process and as such the formal results of TPT can be applied. 

\subsection*{Transition path theory}
With a stochastic process description of the quantum dynamics afforded by the unravelled Lindblad equation, we will use TPT to resolve the salient features of the resultant reactive dynamics. TPT presents a framework for characterizing the reactive dynamical events of a system described by a Markov process. The central quantity in TPT is the transition matrix, which is the integrated infinitesimal generator.\cite{wales2018exploring,sharpe2021nearly} The transition matrix has entries $T_{i,j}$ that indicate the probability for state $i$ to transition to state $j$ during time $\tau$.\cite{noe_schutte}
We consider a Markov model spanned by the energy eigenbasis with transition probabilities informed by Lindblad population calculations. Specifically, we calculated $T_{i, j}$ from density matrix propagation beginning from the reduced density matrix $\rho_{ii}(0)=  \avg{| \phi_i \rangle \langle \phi_i|}$, 
propagated over a short time $\tau$ as
\begin{equation}
\label{Eq:Tij}
T_{i,j} =  [e^{\tau \mathcal{D}} \rho_{ii}(0)]_{jj} \approx \left \{1+\tau \mathcal{D}[ \rho_{ii}(0)] \right \}_{jj} 
\end{equation}
where $\mathcal{D}$ is the superoperator defined in Eq.~\ref{eq:D}. For the systems we consider, $\tau$ can be taken arbitrarily small provided errors due to numerical precision are not encountered, and should not be taken much larger than the characteristic hopping time between energy adjacent states. The dependence of inferred properties on the choice of $\tau$ is explored in App.~
\ref{AppA}.

TPT offers a set of relations to evaluate the commitment probability for transitions between an initial set of states, $A$, referred to  here as the reactant state, and a final set of states, $B$, referred to here as the product state. The commitment probability, or committor, is denoted, $P_{B|A}(i)$, indicating the probability for a system currently in state $i$ to visit the product subset of states $B$ before visiting the reactant subset of states $A$. Similarly the probability, $P_{A|B}(i)$, indicates the probability to visit $A$ before visiting $B$. The reverse committor, $P_{A|B}^*(i)$, is defined as an analogous conditional probability to $P_{A|B}(i)$ but under a time reversed dynamics. In a detailed balance system, the reverse committor $P_{A|B}^*(i)$, is $P_{A|B}^*(i)=P_{A|B}(i)=1- P_{B|A}(i)$ due to microscopic reversibility. The committor functions as an ideal reaction coordinate, as it conveys exactly the progress of a transition from the reactant state to the product state.\cite{onsager1938initial,chodera2011splitting} The collection of states for which $P_{B| A}=1/2$ is then identifiable as a transition state ensemble, as those states have equal likelihood of proceeding to the product state or returning to the reactant state. 

When the full matrix of transition rates is computable, the TPT approach is fast, efficient and rigorously assigns the committor probabilities while simplifying the description of the reaction pathways. 
From the transition matrix, committor probabilities satisfy a backward Kolmogorov equation,\cite{weinan_ve}
\begin{equation}
P_{B|A}(i) - \sum_{j\in I} T_{i,j} P_{B|A}(j) = \sum_{j \in B} T_{i,j} \, ,
\end{equation}
with boundary conditions $P_{B|A}(i) = 0$ for  $i \in A$ and $P_{B | A}(i) = 1$ for $i \in B$, where $I$ is the set of all states not in $A$ or $B$. 

 Once committors have been calculated, the average flux, $f_{i,j}^{A,B}$, between states $i$ and $j$ conditioned on arriving from $A$ and proceeding to $B$ can be found from
\begin{eqnarray}
    f_{i,j}^{A,B} = \pi_i P_{A|B}^*(i)T_{i,j} P_{B|A}(j) \;\;\; i \ne j
\end{eqnarray}
where by construction $ f_{i,i}^{A,B} = 0$ and $\pi_i$ is the steady-state probability of state $i$.\cite{noe_schutte} To find the total reactive flux, $F$, between $A$ and $B$, 
\begin{equation}
    F = \sum_{a \in A, j \notin A} f_{a,j}^{A,B} = \sum_{j \notin B, b \in B} f_{j,b}^{A,B}
\end{equation}
the reactive flux leaving $A$ along all possible connections or the reactive flux arriving in $B$ along all possible connections is summed. 
{ \color{black} The thermal reaction rate in the long time limit is found in TPT from $F$ divided by time, $\tau$,}  
\begin{equation}
\label{TPTrate}
    k_{A,B}=\frac{F}{\tau \pi_A}
\end{equation}
 taking into account a factor of $\pi_A$, where $\pi_A = \sum_i \pi_i P^*_{A|B}(i)$ is the probability that a system in state $i$ is moving from $A$ to $B$. In the following, each state in the Markov model will be either an eigenstate of the system or any generic superposition yet to collapse to an eigenstate.
  In analyzing conical intersection dynamics in the thermal cases, the reactants, $A$, and products, $B$, will comprise only a single eigenstate. For relaxation following vertical excitation, we will be interested in the resultant branching dynamics necessitating further generalization.

The resultant Markov dynamics can be thought of as a graph with the eigenstates as vertices and edge weights given by the reactive flux, $f_{i,j}^+$\cite{noe_schutte}
\begin{equation}
    f_{i,j}^+ = \max [0,f_{i,j}^{A,B}-f_{j,i}^{A,B}]
\end{equation}
which encodes the net traffic over time $\tau$ between states $i$ and $j$. An ensemble of reactive pathways through this graph may be selected by first choosing a pathway along edges from $A$ to $B$, and subtracting from each edge in the pathway the flux, $f_i$ assigned to this pathway, where $f_i$ is equal to the minimum $f_{i,j}^+$ of any $i$, $j$, transition in the pathway. Another pathway may then be assembled from the modified graph in the same manner. Infinitely many ensembles could be assembled in this way. 
An ensemble of particular interest is the max-min flux ensemble as it repeatedly locates the bottleneck of the graph and incorporates this edge into the next pathway.\cite{weinan_ve,noe_schutte} In this scheme, a pathway with the largest possible flux is chosen at each  step. A modified Dijkstra's algorithm, employed repeatedly, can locate these pathways.\cite{dijkstra}

{\color{black}  
Since we define reactants and products based on energy eigenstates, the current method requires that the system has low energy eigenstates that are localized in each well. Such definitions would not work for a symmetric system that results in the lowest energy eigenstate or eigenstates being delocalized, nor for the case in which the energy barrier is so low that significant mixing of diabatic character occurs in the very lowest eigenstates. In the former case, for large barriers and small couplings, the reactive flux would be insensitive to a small perturbation that would break the symmetry providing a potential generalization. In the latter case, a steady-state thermal rate would not likely be definable.}
\section*{Linear vibronic coupling model}

\begin{figure}[t]
\begin{center}
\includegraphics[width=8.5cm]{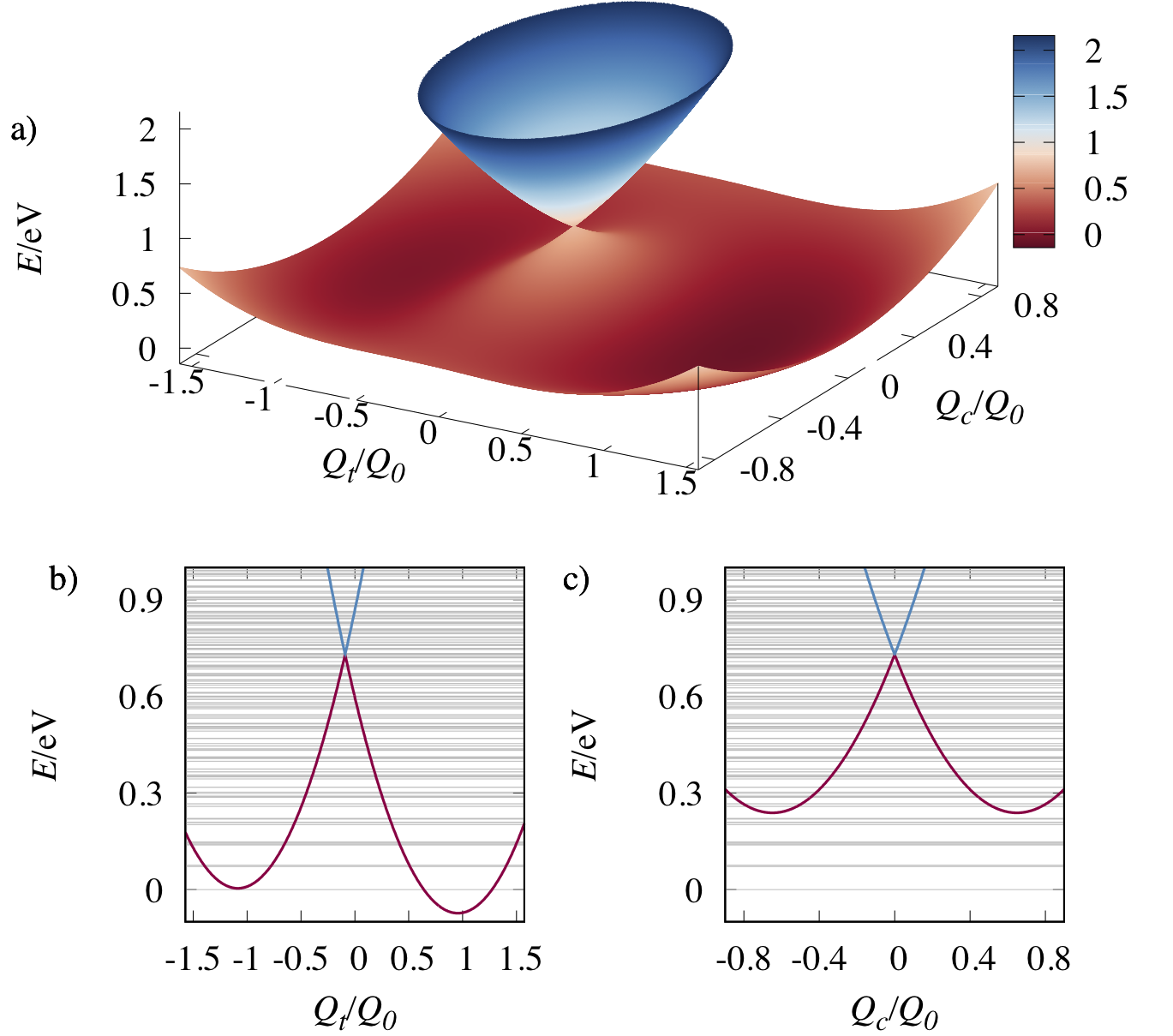}
\caption{
a) Adiabatic surfaces for $\lambda=1.3\lambda_0$ with the color indicating energy.  b) Slice through the conical intersection along $Q_c/Q_0=0$ with adiabatic energy surfaces in red and blue, and eigenstate energies in gray. c) Slice through the conical intersection along $Q_t/Q_0=-0.09$ denoted as in b).
}
\label{ci_plot}
\end{center} 
\end{figure}

In order to test the utility of TPT for nonadiabatic dynamics, we have studied a linear vibronic coupling model of a conical intersection. The linear vibronic model is a multistate, multimode harmonic oscillator model. Here we focus on two modes, a tuning mode, $Q_t$, and a coupling mode, $Q_c$, which define a conical intersection.
The system Hamiltonian is\cite{kouppel1984multimode} 
\begin{equation}
    H_s = \sum_{k=1,2} | \psi_k \rangle h_k \langle \psi_k | + \left(|\psi_1 \rangle \langle \psi_2 | +  | \psi_2 \rangle \langle \psi_1|\right) \lambda Q_c
\end{equation}
where $\psi_k$ denotes the $k$th diabatic state. The diabatic states are coupled through $Q_c$ with strength determined by the diabatic coupling $\lambda$. Each diabatic Hamiltonian, $h_k$, is given by
\begin{equation}
\label{Eq:ham}
    h_k = \frac{1}{2} \sum_{j=c,t} \hbar \omega_j \left\{P_j^2 + Q_j^2 \right\} + E_k + \kappa_k Q_t
\end{equation}
where $P_j$ and $Q_j$ are the momentum and position operators of the coordinate $j$ and $E_k$ is a constant energy added to each diabatic potential. The oscillator frequencies are given by $\omega_j$, and tuning oscillator position displacements by $\kappa_k$. Examples of the resultant adiabatic potential energy surfaces, as well as the energies of lower lying eigenstates, are shown in Fig.~\ref{ci_plot} for the case of large diabatic coupling. 
We consider a sufficiently small range of $\lambda$ such that there is a barrier in the lower adiabatic potential energy surface that supports localized energy eigenstates within each basin.

The baths coupled to $Q_c$ and $Q_t$ consist of an infinite number of harmonic oscillators 
\begin{eqnarray}
    H_B = \sum_\alpha \sum_{j=c,t} \frac{1}{2} \hbar \omega_{\alpha,j} \left(p_{\alpha,j}^2 + q_{\alpha,j}^2 \right)
\end{eqnarray}
where $p_{\alpha,j}$ and $q_{\alpha,j}$ refer to the  momentum and coordinate of the $\alpha$'th harmonic oscillator in the bath coupled to the $j$'th mode in the system.
The system-bath coupling is bilinear such that
\begin{equation}
    V = \left( |\psi_1\rangle \langle \psi_1| + |\psi_2 \rangle \langle \psi_2 | \right) \sum_\alpha \sum_{j=c,t}  c_{\alpha,j} q_{\alpha,j}Q_j  
\end{equation}
where $c_{\alpha,j}$ is the coupling strength of each oscillator.
The coupling of the system to the bath is summarized by a spectral density of Debye form,
\begin{eqnarray}
    J_j(\omega) &=& \sum_\alpha c_{\alpha,j}^2 \delta(\omega - \omega_{\alpha,j}) \nonumber \\ 
    &=&2 \eta \frac{\omega \omega_{b} }{\omega^{2} + \omega_{b}^2}
\end{eqnarray}
where $\omega_b$ is the cut off frequency and $\eta$ is the reorganization energy which is taken to be small.

The specific parameters we have employed for this model are reminiscent of a parameterization for the photoisomerization of pyrazine.\cite{chen_lipeng} However, in order to produce a metastable, double-well structure in the ground adiabatic state, we increased the displacement of both oscillators relative to the parameters previously used and adjusted the relative energies of the diabatic electronic states. See Table \ref{tab1} for specific parameters employed. Of particular importance will be the impact that the diabatic coupling strength $\lambda$ on the reactive dynamics in the vicinity of the conical intersection. The magnitude of $\lambda$ has a dramatic impact on the conical intersection geometry, particularly on the adiabatic potential energy barrier height. The value of $\lambda$ is frequently changed by a multiple throughout this work with $\lambda_0$ being the value found in the unaltered pyrazine model, $\lambda_0 = 0.262 \; e$V. To simplify figures, we define $Q_0$ as the average of $\avg{Q_t}$ of the lowest eigenstates respectively localized in each diabatic state in the $\lambda=1.3\lambda_0$ system.

\begin{figure}[b]
\begin{center}
\includegraphics[width=8.5cm]{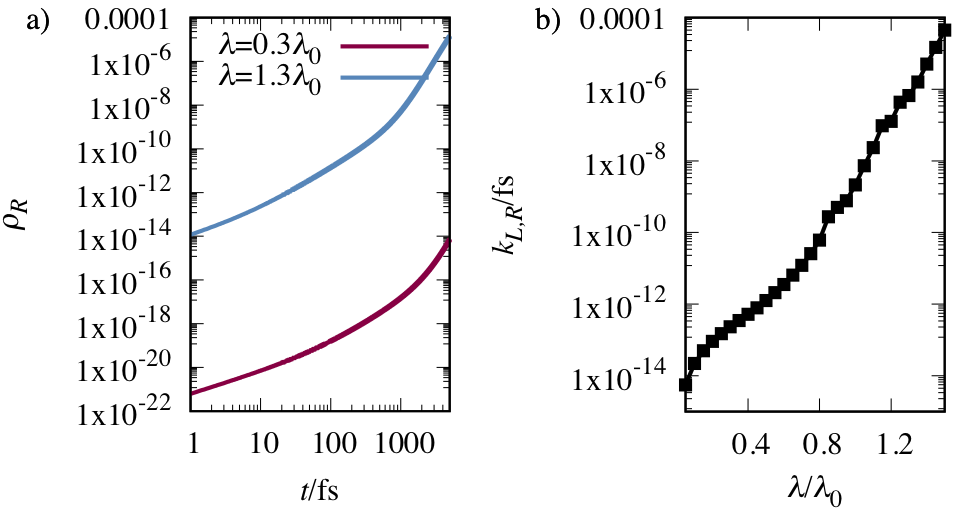}
\caption{a) Population in state $R$, $\rho_R$, as a function of time from density matrix dynamics initialized in $L$ for a low and high $\lambda$ system. b) Rates of population transfer between $L$ and $R$, $k_{L,R}$, as a function of $\lambda$.}
\label{eq_ensemble}
\end{center} 
\end{figure}
\section*{Thermal barrier crossings}
We first consider the transition paths and reactive rates associated with a thermal barrier crossing in the vicinity of a conical intersection. Employing TPT within a Markov model spanned by the energy eigenstates, we abstract away complications associated with defining reactive paths in the presence of both nuclear quantum effects and nonadiabatic effects. Rather than consider reactive paths localized in position space, the paths are defined through a sequence of states localized in energy. In the limit of a high barrier, we expect to recover results from quantum instanton theory.\cite{ranya_ananth,heller2020instanton}

For a thermal transition, the system evolves with a detailed balance dynamics, with an incoherent initial condition. In this case, TPT can be used straightforwardly. For concreteness, we consider transitions between the lowest energy eigenstates localized in each diabatic state, and designate them as $L$ and $R$, with $R$ being the lowest energy eigenstate in the system. The specific states depend on $\lambda$. Microscopic reversibility guarantees that reactive pathways from $L$ to $R$ are direct inverses of reactive pathways from $R$ to $L$, making the direction of barrier crossing under study irrelevant.

\subsection*{Typical behavior}

Figure \ref{eq_ensemble}a shows the population in $R$ where $\rho_R(t)= \avg{|\phi_R \rangle \langle \phi_R | }   $, following initialization of the system in $L$ under density matrix evolution. Two different $\lambda$ values are displayed and although $\rho_R$ increases more quickly with time for the larger $\lambda=1.3\lambda_0$ case, over $1$ ps a negligible population has accumulated in $R$. For the lower coupling, $\lambda=0.3\lambda_0$, the accumulation of population is many orders of magnitude smaller. This slow rise in population reflects the large adiabatic barrier separating $R$ and $L$ with a concurrent small rate constant. The barrier height depends sensitively on the size of $\lambda$, decreasing rapidly as $\lambda$ increases.  Apart from this slow increase in population, there are no other discernible features in the average dynamics. 

The rate constant $k_{L,R}$ for the barrier crossing event from $L$ to $R$ calculated using Eq.~\ref{TPTrate} is shown in Fig \ref{eq_ensemble} b. The rate constant increases monotonically with increasing coupling strength. For very small coupling, the rate constant grows in proportion to $\lambda^2$ as expected from perturbation theory, but the domain of that scaling relationship is small $\lambda/\lambda_0 \ll 1$. 
For larger $\lambda$, the rate increases rapidly with a mechanism that from the mean wave packet propagation is not easily discernible. 

\begin{figure}
\begin{center}
\includegraphics[width=8.5cm]{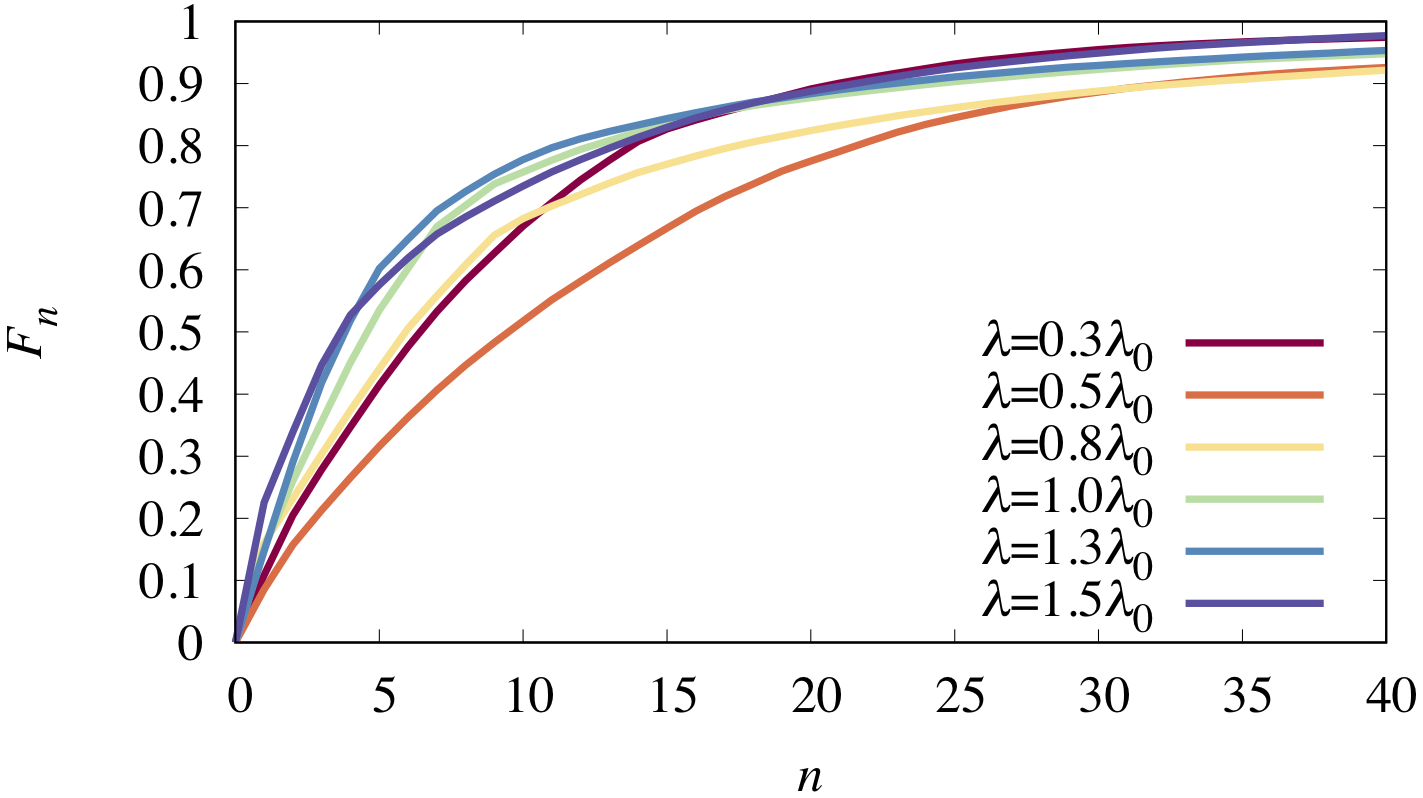}
\caption{The number of thermally calculated pathways, $n$, required to account for a given fraction, $F_n$, of the overall thermal flux between $R$ and $L$ for several values of $\lambda$. 
}
\label{eq_flux}
\end{center} 
\end{figure}

\begin{figure}
\begin{center}
\includegraphics[width=8.5cm]{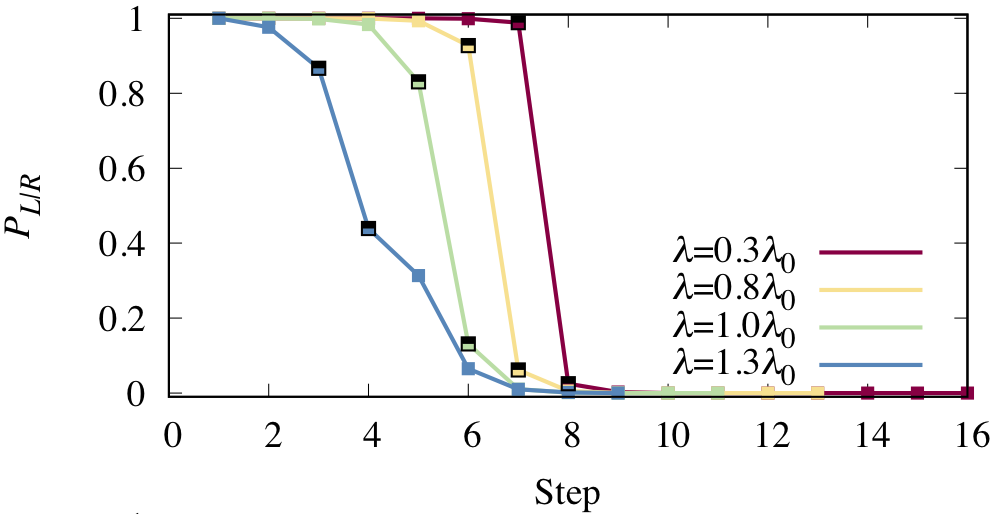}
\caption{Dominant thermal barrier crossing pathways showing the probability to commit to $L$, $P_{L|R}$, for several $\lambda$. Filled symbols denote the transition wavefunctions just prior to and after committing.}
\label{eq_paths}
\end{center} 
\end{figure}
\subsection*{Transition path ensemble}

Each pathway in the TPT max-min flux decomposition of pathways from $R$ to $L$ carries flux $f_i$. These paths represent the transition path ensemble. The diversity of paths, and their corresponding weights, can be understood by rank ordering the paths based on their flux. The cumulative flux fraction accounted for by the first $n$ highest flux pathways in the decomposition
\begin{equation}
F_n = \frac{1}{F}\sum_{i=1}^{n} f_i
\end{equation}
is a direct measure of the number of relevant pathways in the transition path ensemble.\cite{noe_schutte} Figure \ref{eq_flux} shows $F_n$ for several different $\lambda$ values. In this system, likely as a result of the large barrier in the lower adiabatic potential, the majority of the reactive flux, regardless of diabatic coupling strength, is accounted for by the first few pathways. This manifests the approach of the instantonic limit, as the barrier height is large compared to both thermal energy as well as the zero point energy in the tuning mode. As the barrier becomes large, only few paths contribute significant weight to the reactive path ensemble. 
Because in this thermally activated process, there are clearly a few dominant pathways, these can be inspected more closely to discover the most prevalent transition mechanisms at different coupling strengths.

\subsection*{Dominant transition paths}

We inspected some of the prominent transition pathways for mechanistic information. We considered the transition path from $L$ to $R$ with maximum $f_i$ at several $\lambda$ values. Figure \ref{eq_paths} shows the committor, $P_{L|R}(i)$, for each step in the pathways, meaning the probability at each step for the system to return to eigenstate $L$. The principle paths are the most likely sequences of states along a reactive quantum jump trajectory but do not retain direct temporal information. To recover a typical timeseries, these paths would need to be convoluted with the appropriate waiting time distributions at each step. Nevertheless, the sequence of states retain significant mechanistic information. 

At low $\lambda$, the transition pathways involved more quantum jumps with the dominant pathway at $\lambda/\lambda_0=0.3$ requiring 16 jumps compared to 9 for $\lambda/\lambda_0=1.3$. As $\lambda$ increased, the transition eigenstates, the states just prior to and just after committment defined at $P_{L|R}=1/2$, became closer to $P_{L|R}=1/2$, more mixed in terms of diabatic character and less localized in either well. This behavior can be understood, as the states near the conical intersection are nearly degenerate and more  susceptible to delocalization with increasing $\lambda$.  
In all cases, a single quantum jump, interpretable as a barrier crossing event, resulted in a large change in average committor value. The abrupt transitions reflect tunneling contributions to the reaction paths, as quantum mechanically the system need not actually pass through intermediate states to move from one side of the barrier to another. 

\begin{figure}[t]
\begin{center}
\includegraphics[width=8.5cm]{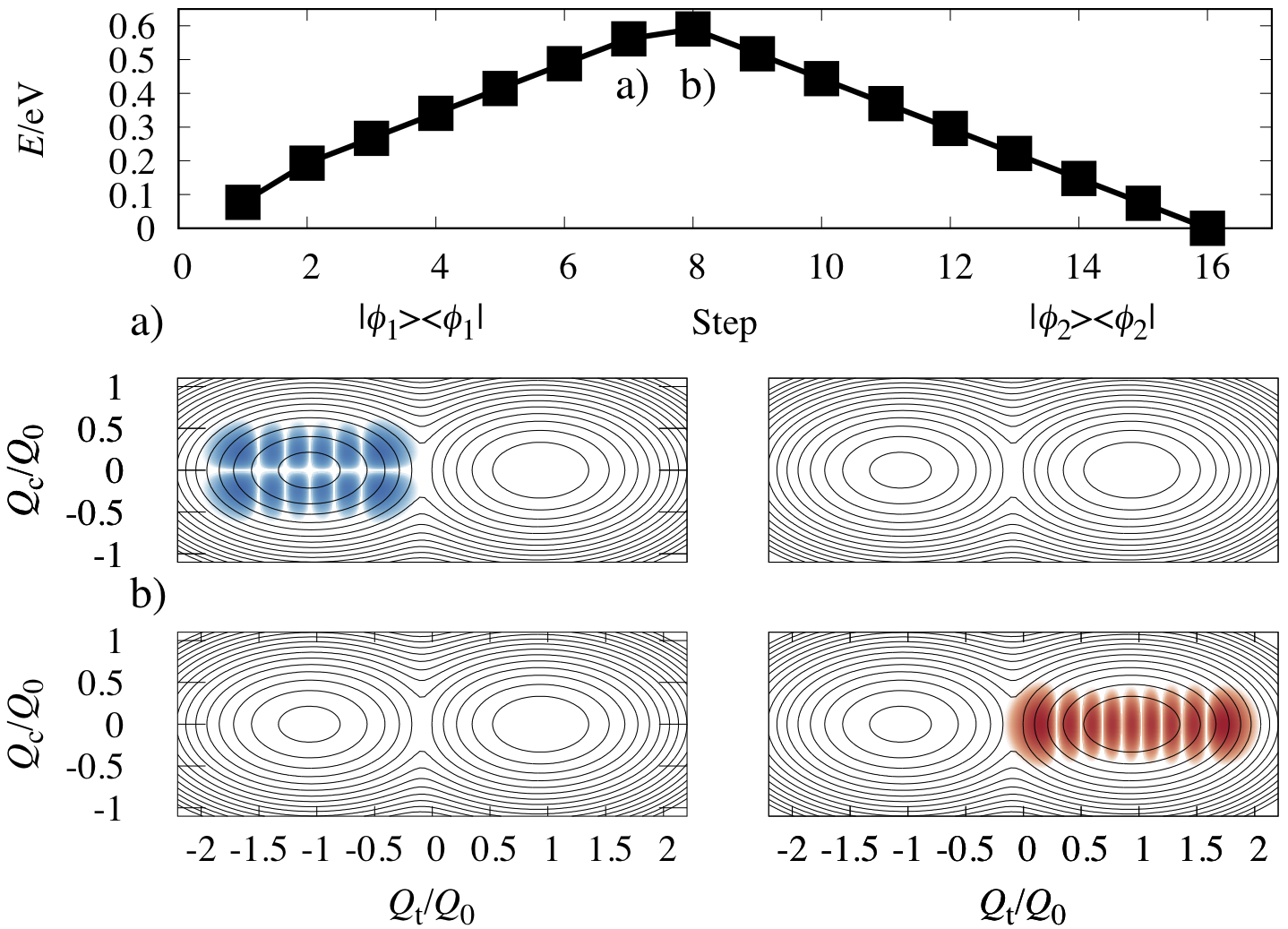}
\caption{(top) The energy at each eigenstate along the highest $f_i$ thermal transition pathway between $L$ and $R$ with $\lambda=0.3\lambda_0$. (bottom) Transition wavefunction densities for the state before $P_{R|L}$ exceeds $1/2$ (a) and after $P_{R|L}$ exceeds $1/2$ (b) plotted on a logscale. Red indicates density in diabatic state 2 and blue indicates density in diabatic state 1. Superimposed on the transition wavefunctions are the lower adiabatic potentials with contours placed at intervals of $0.136\; \mathrm{eV}$.}
\label{eq_03}
\end{center} 
\end{figure}

\begin{figure}[t]
\begin{center}
\includegraphics[width=8.5cm]{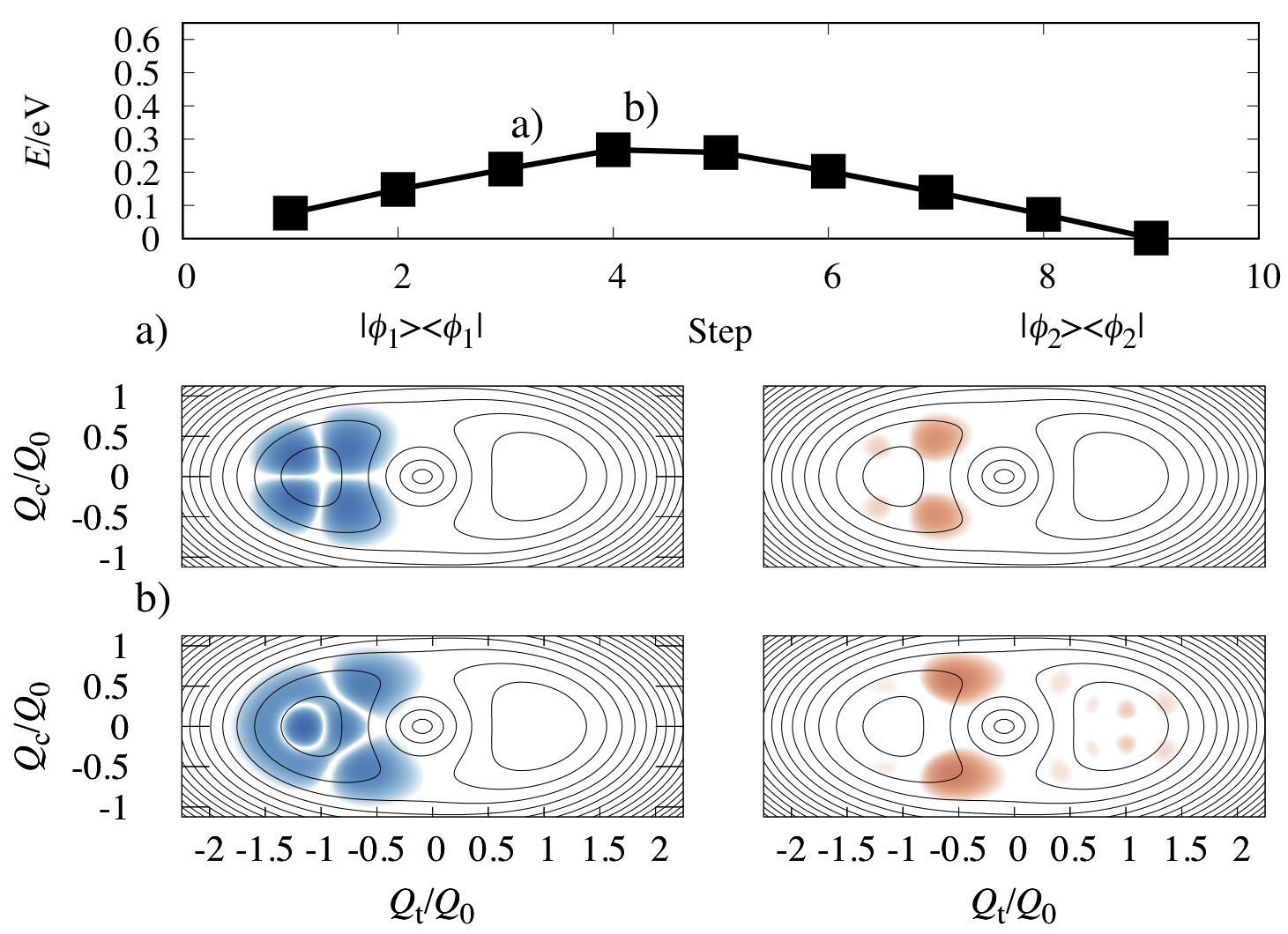}
\caption{(top) The energy at each eigenstate along the highest $f_i$ thermal transition pathway between $L$ and $R$ with $\lambda=1.3\lambda_0$. (bottom) Transition wavefunction densities for the state before $P_{R|L}$ exceeds $1/2$ (a) and after $P_{R|L}$ exceeds $1/2$ (b) plotted on a logscale. Red indicates density in diabatic state 2 and blue indicates density in diabatic state 1. Superimposed on the transition wavefunctions are the lower adiabatic potentials with contours placed at intervals of $0.136\; \mathrm{eV}$.}
\label{eq_13}
\end{center} 
\end{figure}

We inspected the wavefunctions along the barrier crossing pathways and found that low and high $\lambda$ resulted in significantly different behaviors. Figures \ref{eq_03} and \ref{eq_13} show the energy, $E$, as a function of step along the dominant reaction pathways for different $\lambda$ values as well as the transition eigenstate wavefunctions projected onto each of the diabatic electronic states. 
In the low $\lambda$ limit in Fig. \ref{eq_03}, both transition eigenstate wavefunctions are similar to harmonic oscillator wavefunctions. They show negligible overlap in density and their energies are well below the barrier height, which is approximately at $0.7 \; \mathrm{eV}$, although the energies of the transition eigenstates are still quite high. This indicates a deep tunneling barrier crossing mechanism and is consistent with sharp changes in position and diabatic character at the transition.

In the high $\lambda$ limit in Fig. \ref{eq_13}, the transition pathway was relatively short, passed through eigenstates with energies nearly equal to the height of the barrier at approximately $0.24 \; \mathrm{eV}$ and involved transition eigenstate wavefunctions which did not resemble harmonic oscillator wavefunctions but rather showed significant mixing of both diabatic states. Transition wavefunctions are far less localized in either well with significant density at the barrier of the conical intersection. This indicates barrier crossing takes place by going around the conical intersection rather than by tunneling through it. These pathways are representative of the behavior of the ensemble, although some prominent tunneling pathways remain.

The dominant paths found here are conceptually similar to the dominant pathways produced by the path integral instanton method.\cite{ranya_ananth,cao_voth1995}
{\color{black}  
Previous work of Ranya and Ananth and also Cao and Voth using nonadiabatic instanton theory to analyze the reactive behavior in an avoided crossing and spin-boson model noted that, at smaller diabatic coupling, sharper transitions between diabatic states are observed \cite{cao_voth1995,ranya_ananth}. Cao and Voth also observe an increase in electron transfer rate with increasing diabatic coupling strength and note two regimes, a Golden Rule regime and an Adiabatic limit regime, the later of which results in a much stronger dependence of rate upon diabatic coupling strength. These trends are consistent with our observations.}\cite{cao_voth1995}

\begin{figure}
\begin{center}
\includegraphics[width=8.5cm]{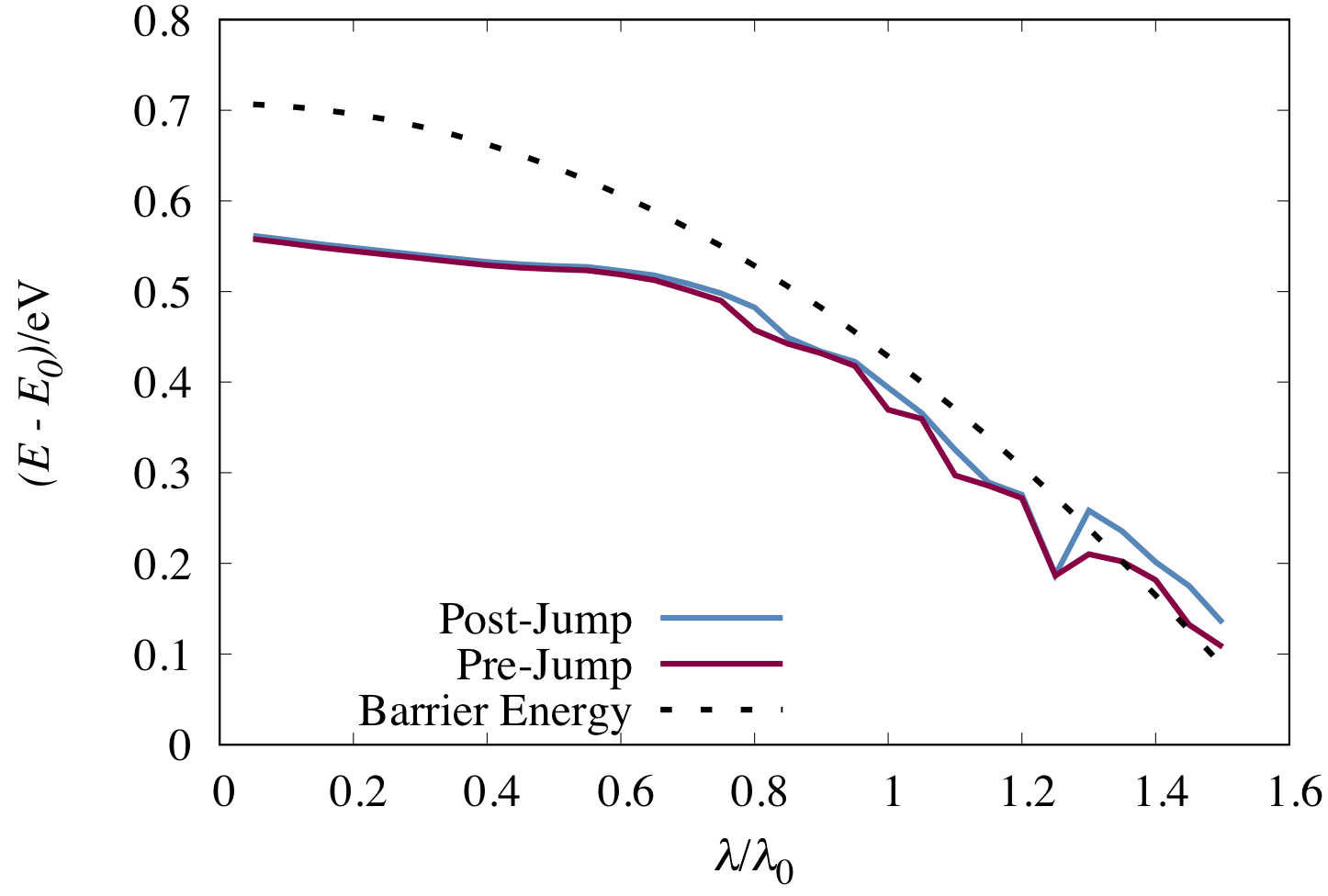}
\caption{Average energy relative to $E_0$, the energy of the lowest eigenstate for the given system, of the transition eigenstates, post-jump and pre-jump, and barrier height during the thermal barrier crossing as as a function of $\lambda/\lambda_0$ averaged over the transition path ensemble.}
\label{eq_enbar}
\end{center} 
\end{figure}

The trend observed in the representative pathways suggests that tunneling is the main barrier crossing mechanism at low $\lambda$ and traversing around the conical intersection is the main barrier crossing mechanism at high $\lambda$. This interpretation is evident in Fig. \ref{eq_enbar} which displays the energy of the transition eigenstates versus $\lambda$ relative to the height of the bare lower adiabatic potential barrier. {\color{black}   Note that all thermal data was generated from exact density matrix propagation so there are no statistical errors. The sharp features in Fig. \ref{eq_enbar} arise from accidental degeneracies between low-lying energy eigenvalues.}
 
At low coupling, the average energy has little dependence on $\lambda$ and is found far below the barrier. These eigenstates below the barrier display little mixing between electronic states and are highly localized in one well or the other. Deep tunneling is the only feasible reaction pathway when the barrier is so high. This is a very slow mechanism, as demonstrated in Fig. \ref{eq_ensemble} where low coupling results in slow population transfer between $L$ and $R$ and a low $k_{L,R}$. As $\lambda$ increases, the average energy of transition eigenstates for the ensemble approaches that of the barrier, leading to more delocalization and shallower tunneling mechanisms. The transition eigenstate energies surpass the barrier at the highest coupling strengths only. The barrier crossing proceeds in this case by going around the conical intersection, a significantly faster mechanism that results in more mixing of electronic character in the transition eigenstates.

{\color{black} The transition between the small and large $\lambda$ limits corresponds to an end of the Golden rule regime where the rate scales like $\lambda^2$. Such a regime approximately requires that $\lambda$ is small relative to the frequency of the coupling coordinate, though a full accounting on the non-Condon effects at finite temperature complicates a simple analytical picture.} \cite{heller2020instanton,izmaylov2011nonequilibrium}  By constructing transition matrix elements directly from the eigenstate wavefunctions, we maintain any relevant phase relationships upon transitions around the conical intersection such as those due to the generation of a geometric phase.\cite{ryabinkin2013geometric} {\color{black} Employing the procedure in Ref. \onlinecite{izmaylov2016diabatic} we find that geometric phase effects in the large $\lambda$ are small, suppressing the rate by only around 1$\%$ for the $\lambda=1.3\lambda_0$ case}.

\section*{Relaxation following vertical excitations}
Although the equilibrium behavior of a conical intersection system is relevant to thermal isomerizations of photoswitches, it is typically the relaxation following vertical excitation that is of more interest as the energy for spontaneous isomerization is large, and much more readily accessed through photoexcitation than thermal fluctuation. Upon photoexcitation a barrier crossing is no longer rare, though during relaxation unlikely events may cause trajectories to favor one potential product state over another. The commitment of a relaxing trajectory to one well or another of the conical intersection can thus be a non-equilibrium rare event. 
While the bulk of the machinery of TPT is not applicable, some ideas for trajectory analysis can be applied to distill the complicated system dynamics. The committors are independent of the initialization of the system, but density matrix propagation information alone is no longer sufficient to analyze relaxation from a vertical excitation because the initial distribution is not stationary. Because the dynamics are Markovian, the transition matrix still provides complete information about the subsequent relaxation, but the uncollapsed wavefunction following vertical excitation requires special consideration. 

To study relaxation from a vertical excitation, we consider an initial condition generated from a projection of the lowest energy vibrational eigenstate of a harmonic oscillator located at $Q_i/Q_0=0$ for $i=c,t$ with the same frequencies as employed in Eq. ~\ref{Eq:ham}, into diabatic state $|\phi_1\rangle$. Such a condition has been studied previously in the context of the pyrazine model we have adapted\cite{chen_lipeng} and is meant to approximate the excitation from the ground state into a manifold of excited states.  Through application of suitably generalized TPT methods to the vertical relaxation case, we are able to determine how relaxing trajectories subsequently commit to one well or another, employing our previous definitions of the $R$ and $L$ states.

\subsection*{Typical behavior}
\begin{figure}[t]
\begin{center}
\includegraphics[width=8.5cm]{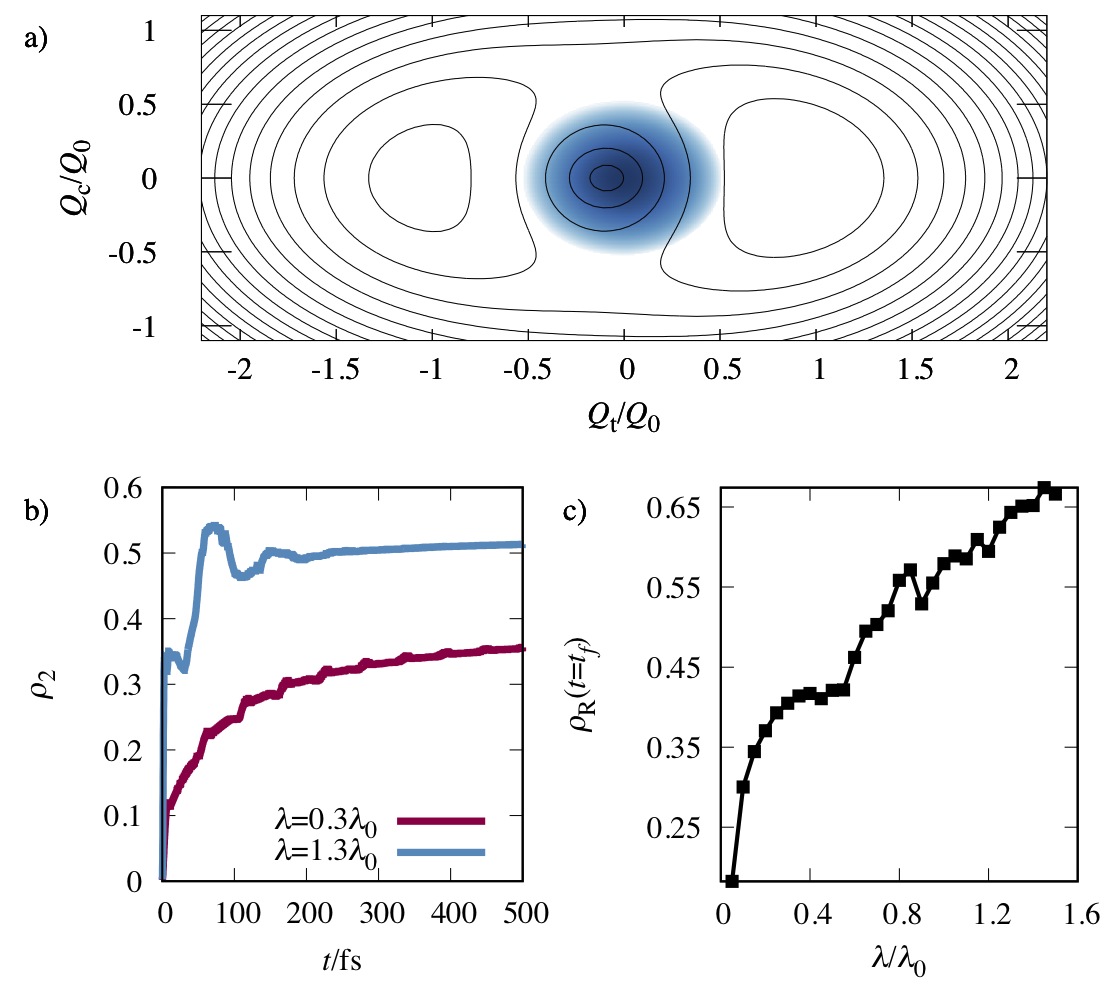}
\caption{
{\color{black}  
a) Initial, vertically excited wavepacket density on a log scale with all density being found in diabatic state 1 depicted in blue. b) Average diabatic state $2$ population, $\rho_2$, for two extreme coupling strengths following vertical excitation into diabatic state 1. c) Photoyield of state $R$, $\rho_R(t=t_f)$ following vertical excitation into state $1$ at various $\lambda$. 
}
}
\label{avg_vert}
\end{center} 
\end{figure}

 We observe the typical behavior of the system in Fig. \ref{avg_vert} b) which displays the average relaxation from direct density matrix propagation following vertical excitation for two $\lambda$ values. The population in diabatic electronic state $2$, $\rho_2 =\avg{| \psi_2 \rangle \langle \psi_2 | }$, shows comparatively quick relaxation towards its equilibrium population with the fast oscillations indicative of $Q_t$ vibrations. Significantly larger but more quickly damped oscillations are seen for the relaxation at higher coupling strength, $\lambda=1.3\lambda_0$. The higher coupling strength also results in a larger proportion of population relaxing into diabatic well $2$. Both the increase in speed of relaxation and increase in proportion of trajectories which change diabatic state during relaxation with increasing $\lambda$ are expected from perturbation theory. 

We also study the average photoyield, computed using a first passage procedure, whereby states $R$ and $L$ are fixed as absorbing boundary conditions. In practice this was computed using 10,000 Lindblad jump trajectories initialized in the
{\color{black}  
vertically excited state shown for $\lambda=1.3\lambda_0$ in Fig. \ref{avg_vert} a. The slight asymmetry in the initial wavepacket is a consequence of basis truncation.
}
The photoyield, $\rho_R(t=t_f)$, where $t_f$ is the time at which a trajectory first reaches either $R$ or $L$, is the fraction of trajectories that end in $R$ without visiting $L$. Figure \ref{avg_vert} c) displays the photoyield as a function of $\lambda$. Generally, we find that the photoyield increases with increasing $\lambda$ although the trend is noisy. {\color{black} Standard deviations estimated from block averaging are on the order of 0.005. Non-monotonic behavior arises from accidental degeneracies and the change of mechanistic regime.} The plateaus of $\rho_2$ in Fig. \ref{avg_vert} b) and the $\rho_R$ values for the same $\lambda$ in Fig. \ref{avg_vert} c) are not exactly the same. Not only do these values represent different projections of the wavefunction, but $\rho_2$ as a function of time from density matrix calculations can include recrossing events between eigenstates $R$ and $L$ as well as the contribution of uncollapsed wavefunctions. As in the thermal case, little detail of the underlying dynamics is forthcoming from the density matrix propagation alone.

\subsection*{Generalization of TPT}
In order to apply TPT to a vertically excited initial condition and to study the subsequent branching of the relaxing trajectories, we generalize the TPT framework previously presented. To study the branching of a relaxing stochastic process into distinct basins, we define commitor functions slightly more generally. Rather than define forward and backward commitors as compliments of each other, we consider trajectory ensembles in which the system relaxes from some initial state $C$ into a specific product state $L$ or $R$ without visiting the opposing product state. This implies that the forward committor for product state $R$ is defined as the probability of reaching $R$ from state $j$ before returning to $C$ or reaching $L$, denoted as $P_{R|CL}(j)$. The corresponding backward committor is defined as the probability in the time reversed dynamics of reaching the initial state $C$ before states $R$ or $L$, denoted for state $i$ as $P_{C|RL}^*(i)$. Under such conditioning the reactive flux into state $R$ is
\begin{equation}
f_{i,j}^{C,R} = P_{C|RL}^*(i)\pi_i T_{i,j}P_{R|CL}(j)
\end{equation}
where an equivalent formulation can be made for trajectories bound for $L$. 

In addition, the initial condition upon vertical excitation is in general a coherent superposition that is irreversibly decohered by the action of the bath. Prior to decoherence there is not a simple classical means of describing the state of the system from which to evaluate a transition probability using Eq.~\ref{Eq:Tij}. The state of the system is uncertain with delicate phase relationships between the energy eigenstates. Determining the likelihood of any member of the superposition would require a projective measurement and thus loss of the superposition. However, immediately after the state has spontaneously decohered, the system can be described by a classical probability distribution on the energy eigenstates, albeit one that is not Boltzmann distributed and thus not stationary under the Lindblad operator. 

In order to apply TPT, we thus restrict our attention to the Markovian jump dynamics following decoherence and employ an empirical initial distribution generated by evaluating the collapse probabilities into each eigenstate from the initial coherent state conditioned on the final destination of the trajectory. Since the wavefunction collapse is statistical, with significant collapse probabilities into several different eigenstates, it is necessary to scale the reactive flux $f_{i,j}^{C,R}$ by its contribution to the total reactive path ensemble. The contribution of the flux from a specific initial collapsed state $C$ is proportional to the probability for the wavefunction to collapse into eigenstate $C$, denoted $\Pi(C)$, and normalized by the total reactive flux, $F^{C,R}=\sum_{j\ne C,L} f_{C,j}^{C,R}$
for eigenstate $C$. With this weight, the reactive flux is
\begin{equation}
    f_{i,j}^R =  \sum_{C} f_{i,j}^{C,R} \frac{\Pi(C)}{F^{C,R}},
\end{equation}
in the conditioned ensemble bound for $R$. The probability for the wavefunction to collapse into $C$ could be calculated from a conditioned ensemble of Lindblad trajectories. Then TPT could be performed for each eigenstate $C$ as detailed above to determine $f_{i,j}^R$. Due to the exponential factors in $\pi_i$, solving for $f_{i,j}^{C,R}$ is very numerically unstable. Instead of evaluating them and then constructing the conditioned fluxes directly, we sampled trajectories in conditioned ensembles and estimated $f_{i,j}^R$ by counting how many jumps are made between each pair of eigenstates $i$ and $j$. The same trajectory ensemble used to compute the photoyields was used for this purpose.

\subsection*{Transition path ensemble}

\begin{figure}
\begin{center}
\includegraphics[width=8.5cm]{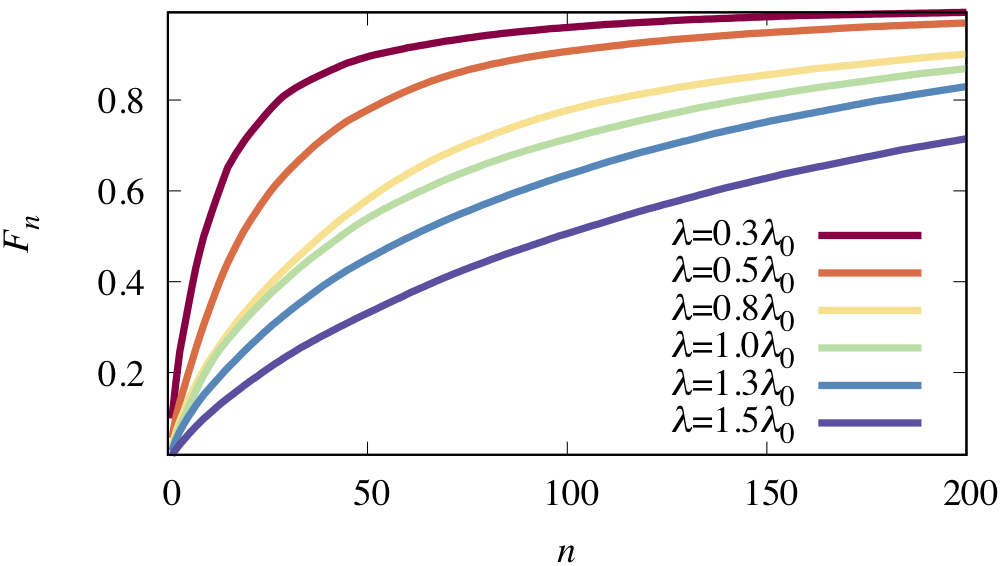}
\caption{
The number of vertical relaxation pathways, $n$, required to account for a given fraction, $F_n$, of the overall flux into $L$ for several values of $\lambda$. }
\label{coll_t}
\end{center} 
\end{figure}

We trace dominant pathways to $R$ through a graph assembled from the ensemble of trajectories ending in $R$ with the edge weights given by $f_{i,j}^{R+}=\max \left [ 0, f_{i,j}^R- f_{j,i}^R \right ]$ then perform the same procedure for trajectories ending in $L$. {\color{black} Note that the rare case in which the first eigenstate on the pathway is also the destination eigenstate is handled in exactly the same way as all other cases and results in a path consisting of a single jump.} Generally, the ensemble of reactive trajectories is much broader than that observed in the thermal reaction. 
Figure \ref{coll_t} shows the cumulative flux accounted for by the first $n$ highest flux pathways in the conditioned ensemble. As the diabatic coupling strength increases, $F_n$ as a function of $n$ flattens out, meaning more pathways are required to account for a given total flux. Ten trajectories are sufficient to account for half of the total flux for the weakest coupling, compared to one-hundred for the strongest coupling. This indicates that at lower coupling strength the variety of pathways available for relaxation is more limited. 

The larger number of pathways at higher coupling is manifested in the much higher number of jumps typically observed during relaxation. The number of inter-eigenstate jumps per trajectory is broader with a higher average for the higher $\lambda$, with a typical number of jumps of 70 for $\lambda= 1.3 \lambda_0$ compared to 10 for $\lambda= 0.3 \lambda_0$. As the coupling to the bath is the same, the larger number of jumps reflects the denser energy eigenstate structure for large $\lambda$, where far more favorably sized energy gaps are accessible. 

\subsection*{Characteristic transition pathways}

\begin{figure}[b]
\begin{center}
\includegraphics[width=8.5cm]{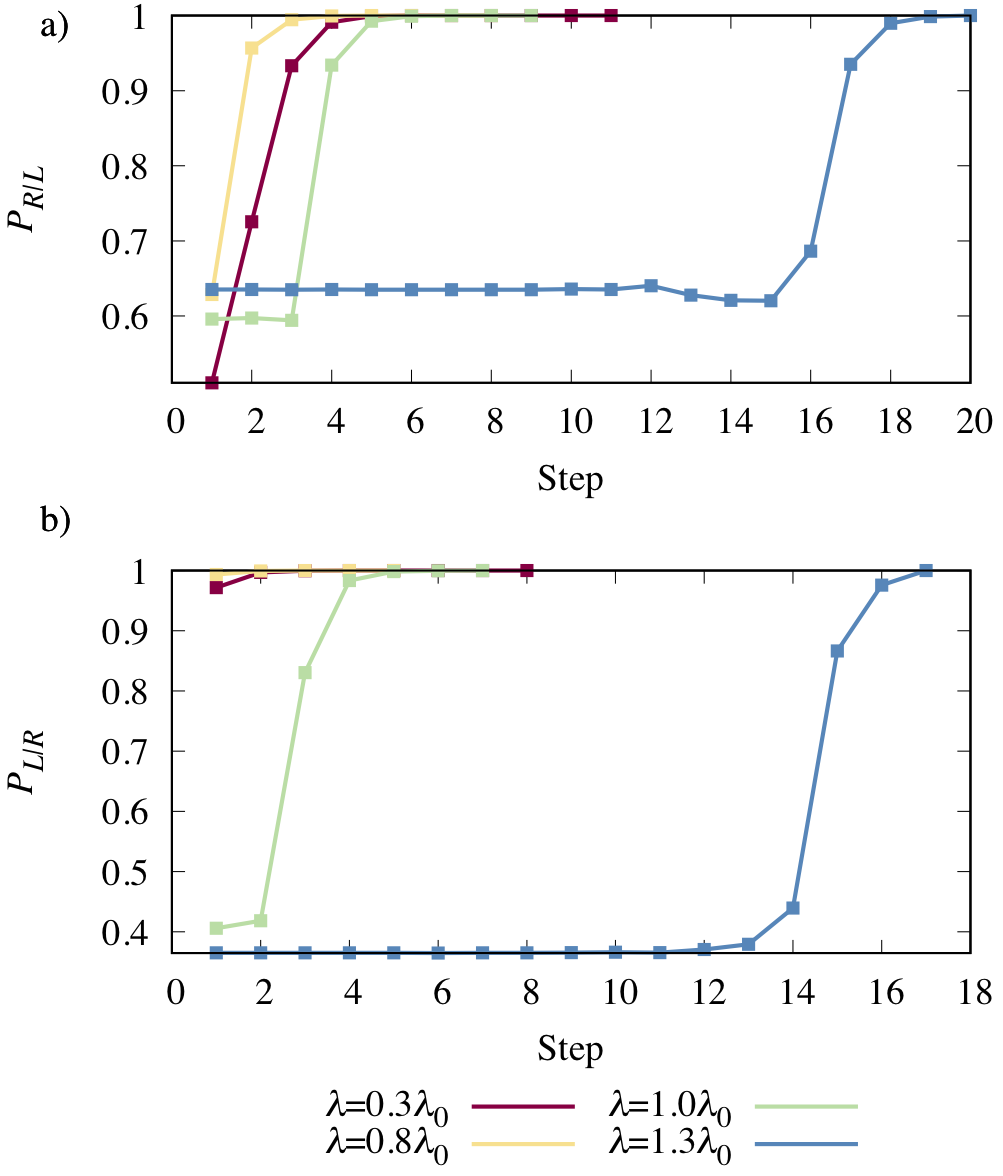}
\caption{Dominant vertical relaxation pathways ending in $R$ and $L$ along with their respective committors, $P_{R|L}$ (a) and $P_{L|R}$ (b).}
\label{vert_paths}
\end{center} 
\end{figure}

While the distribution of reactive trajectories upon vertical excitation is broad, we can nevertheless glean information about the mechanism of branching by considering typical transition paths. Figure \ref{vert_paths} a) shows $P_{R| L }$ along dominant relaxation pathways to eigenstate $R$ following vertical excitation for several $\lambda$ values. Figure \ref{vert_paths} b) shows the compliment $P_{L | R }$ along dominant relaxation pathways to eigenstate $L$.  Immediately upon dephasing to an eigenstate, low $\lambda$ trajectories have committors above $1/2$. Low $\lambda$ trajectories bound for $L$ have $P_{L|R}$ approaching 1.0, indicating that dephasing prior to collapse plays a critical role in determining the outcome of these trajectories even though the state to which they initially collapse  is still very high in energy. In the case of higher $\lambda$, collapse to a moderate $P_{L|R }$ or $P_{R|L }$ eigenstate is followed by jumps that do not alter the committor value, then an abrupt ascent towards a committor value of 1.0. For the strongest coupling case, this is instantonic, superficially similar to the equilibrium case. For high $\lambda$, dephasing seems to be far less important and a set of critical jumps much later decides a trajectory's outcome. More insight into the nature of these jumps can be extracted from closer inspection of the energies and wavefunctions along the relaxation pathways.

\begin{figure*}
\includegraphics[width=17cm]{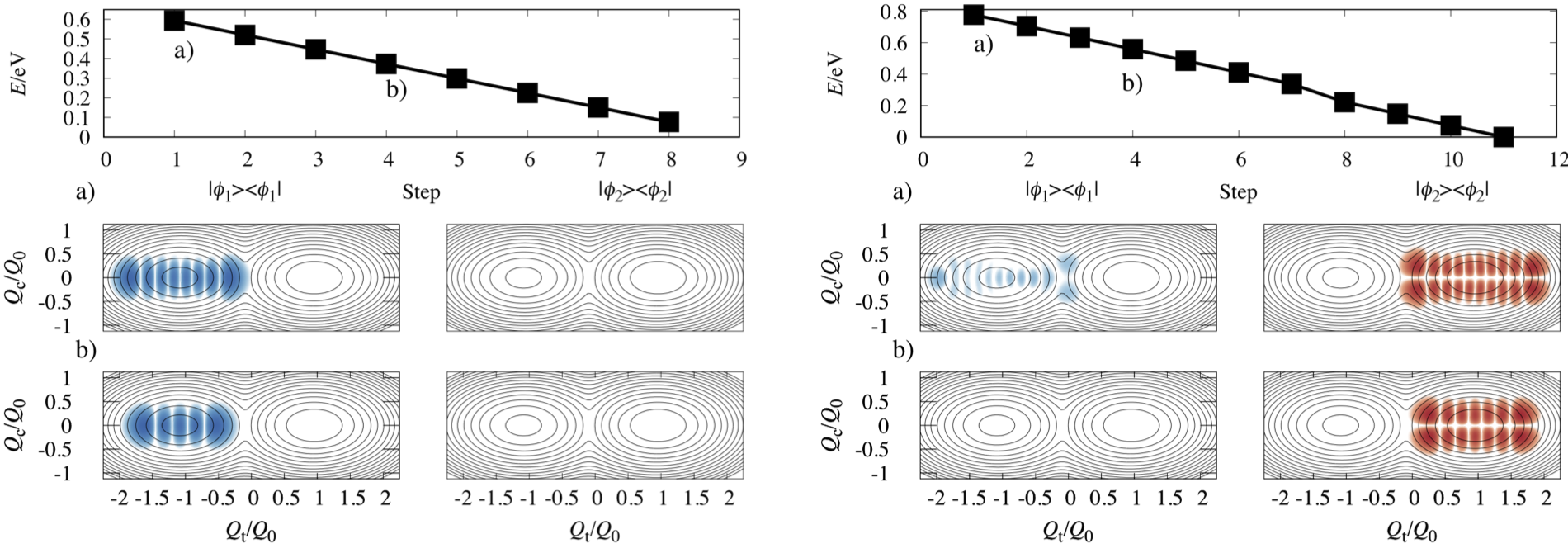}
\caption{ (top) The energy at each eigenstate along the dominant relaxation pathway into $L$ (left) and $R$ (right) when $\lambda=0.3\lambda_0$. (bottom)  The wavefunction density just after collapse into an eigenstate (a) and the wavefunction density following some relaxation (b) plotted on a log scale. All wavefunctions are committed. Blue indicates density in diabatic state 1 and red indicates density in diabatic state 2. Superimposed on the wavefunctions are the lower adiabatic potentials with contours placed at intervals of $0.136\; \mathrm{eV}$.} 
\label{vert_03_A}
\end{figure*}

\begin{figure*}
\includegraphics[width=17cm]{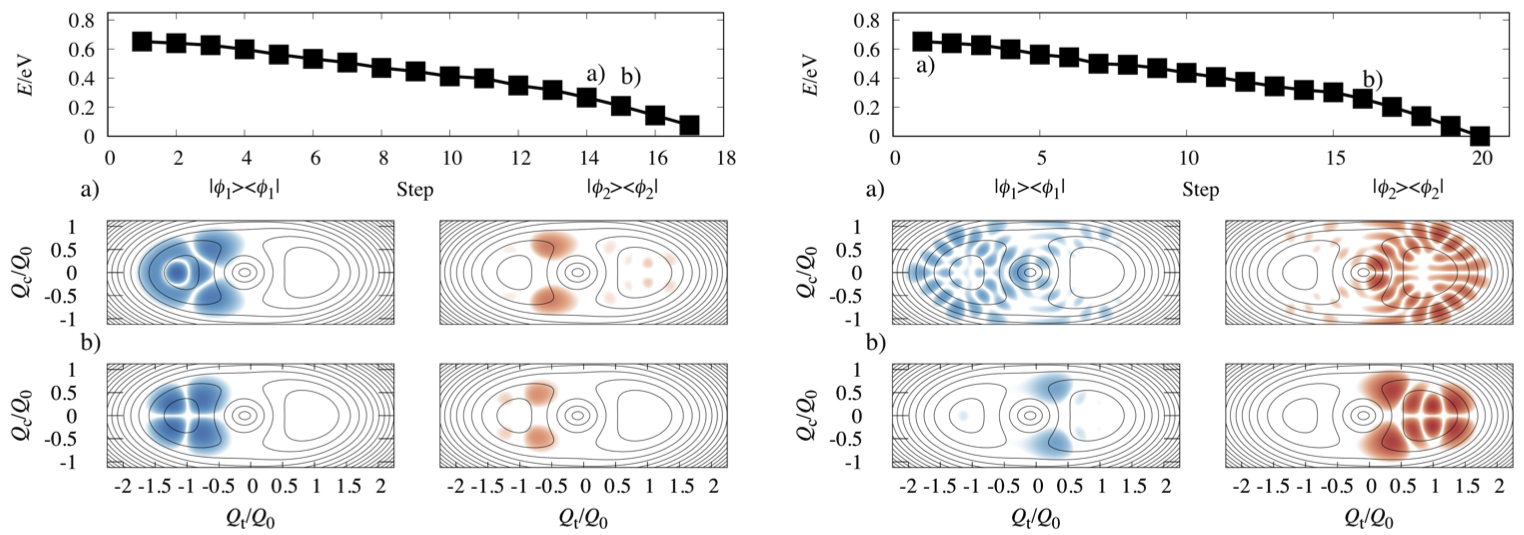}
\caption{(top) The energy at each eigenstate along the dominant relaxation pathway into $L$ (left) and $R$ (right) when $\lambda=1.3\lambda_0$. (bottom) Transition wavefunction densities for the state before $P_{L|R}$ exceeds $1/2$ (a, left) and after $P_{L|R}$ exceeds $1/2$ (b, left) plotted on a logscale. The wavefunction just after collapse into an eigenstate (a, right) and the wavefunction density following some relaxation (b, right).   Blue indicates density in diabatic state 1 and red indicates density in diabatic state 2. Superimposed on the wavefunctions are the lower adiabatic potentials with contours placed at intervals of $0.136\; \mathrm{eV}$.}
\label{vert_13_A}
\end{figure*}

Figures \ref{vert_03_A} and \ref{vert_13_A} show the relaxation pathways to $L$ and $R$ following vertical excitation for weak and strong coupling strengths. In each, we show the energy of the eigenstates along the most likely pathway as well as the wavefunctions projected onto each of the diabatic electronic states for wavefunctions at collapse and after some relaxation, save in the sole case where committor eigenstates can be defined, in which case these states are shown. 

Consider the example trajectories from $\lambda=0.3\lambda_0$. The decision to commit to $R$ or $L$ is made during dephasing with the trajectory committed when it collapses into a single well. This is evident in Fig. ~\ref{vert_03_A} from the harmonic oscillator-like wavefunction centered in well 1 for the trajectory ending in state $L$. The trajectory bound for $R$ is likewise committed and mostly localized to a single well upon collapse, although it does have a small amount of diabatic state $1$ character which quickly disappears as relaxation progresses. 

At large coupling, $\lambda=1.3\lambda_0$, collapses to eigenstates of high energy relative to the barrier are common. As most high energy eigenstates are committed to $R$ in this system, most trajectories, including the examples of trajectories relaxing into $L$ and $R$, collapse into eigenstates for which $P_{R| L }(i)$ exceeds $1/2$. The states into which these pathways collapse are highly delocalized, as evident in Fig.~\ref{vert_13_A}.  It is only after much energy loss through many quantum jumps that the wavefunctions settle into one well or another. At large $\lambda$, commitment to $L$ occurs only after the eigenstate energy becomes comparable to the barrier height. The transition eigenstates in the left panels of Fig. \ref{vert_13_A} have comparable energy to the barrier and have significant density in both diabatic states, but the state for which $P_{L|R }$ surpasses 1/2 has less density in the vicinity of the barrier and is more localized in the metastable well to which it has committed. 

\begin{figure}
\begin{center}
\includegraphics[width=8.5cm]{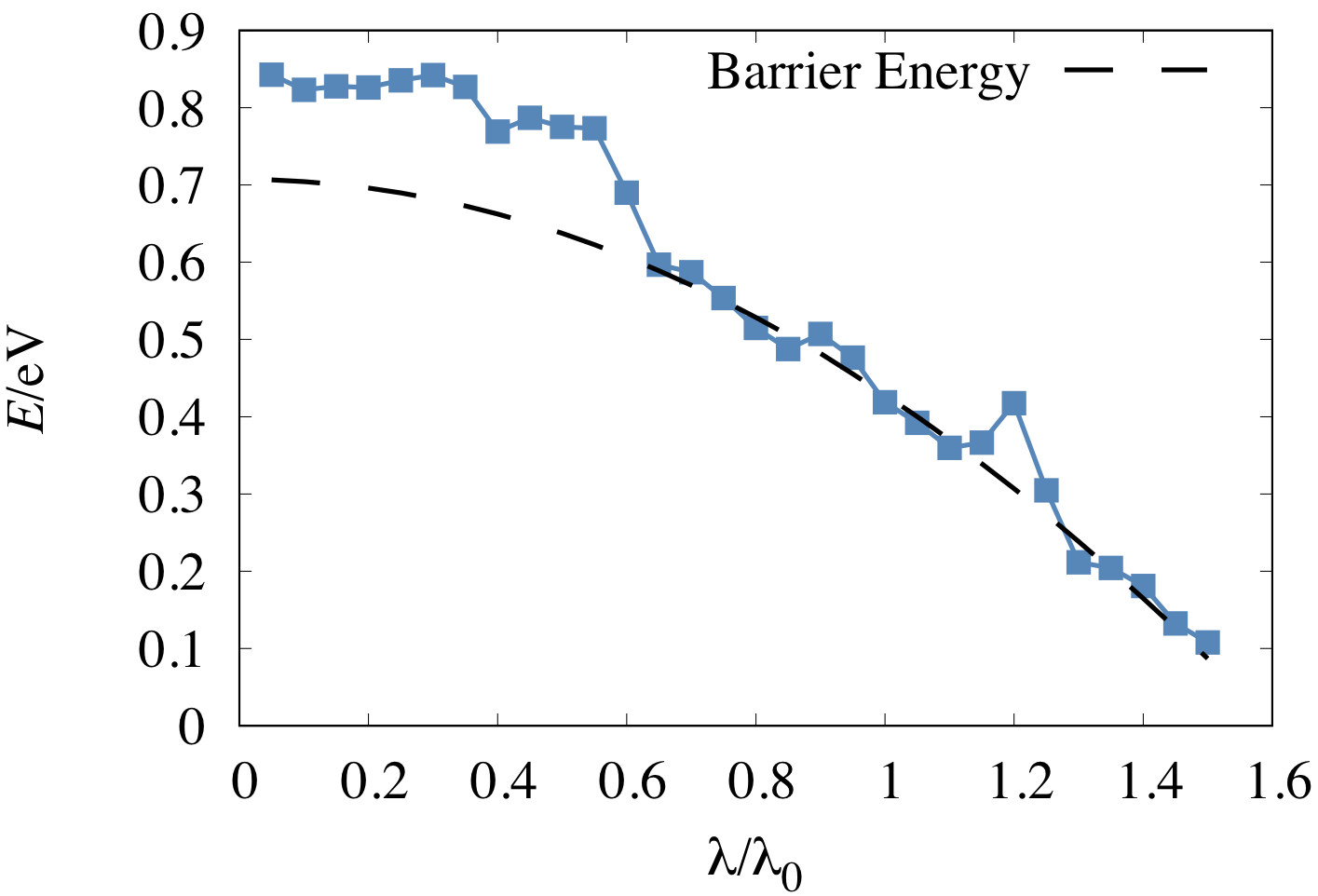}
\caption{Energy relative to the lowest energy eigenstate immediately following the jump at which the trajectory in the ensemble conditioned to end in $L$ first visits an eigenstate for which the $P_{L | R C} \geq 1/2$ and the barrier height for comparison.}
\label{vert_en}
\end{center} 
\end{figure}

Because the wavefunction pre-committor jump may be uncollapsed, statistics about the energy of the state just before the committor surpasses 1/2 could not be reliably collected. However, the post-committor jump eigenstate was available for inspection and its energy for trajectories bound for state $L$ compared with the barrier height at various $\lambda$ is displayed in Fig. \ref{vert_en}.
{\color{black}  
The largest standard deviations estimated from block averaging are on the order of one percent of the average. Accidental degeneracies and the complex interplay of different relaxation pathways leads to the non-monotonic behavior.}
As in the thermal case, there are two characteristic regimes, with a smooth crossover between them. At low coupling, the post-committor eigenstate has energy significantly above the barrier on average. At higher coupling, the post-committor eigenstate has energy comparable to the barrier, supporting the trend shown in the example trajectories where low $\lambda$ systems commit during dephasing and at high energies whereas high $\lambda$ systems commit following dephasing and dissipation until the wavefunction's energy is comparable to that of the barrier. Comparing Fig. \ref{vert_en} with the photoyields in Fig. \ref{avg_vert} b), a lower photoyield is associated with commitment by dephasing, with an apparent regime change at approximately $\lambda/\lambda_0=0.6$. Dephasing commitment provides fewer opportunities to change diabatic states compared to dissipative commitment, a trend reinforced by the larger variety of relaxation pathways observed for higher $\lambda$ in Fig. \ref{vert_paths}.

Lindblad dynamics naturally separates out the effects of the dephasing operator from dissipative operators. Pure dephasing effects have been studied extensively\cite{unruh,skinner_hsu,reina_quiroga,chen_lo,duan_guo,alireza_wang} with studies exploring dephasing and decoherence effects in many different systems. \cite{schlosshauer,naesby_suhr,schlosshauer_hines,ambegaokar,goletz_grossman,banerjee} The importance of dephasing and decoherence to dynamics in more complicated systems such as excitons and conical intersections has been demonstrated. \cite{xiong_chen,heller_joswig} By inspecting the dominant relaxation pathways following vertical excitation, we observe the effects from dephasing separately from dissipation and determine that the action of the dephasing operator decides the fates of trajectories when $\lambda$ is small. {\color{black} The importance of dephasing effects is not surprising as E. R. Heller et. al. recently studied fewest switches surface hopping trajectories} through a conical intersection with various decoherence corrections and determined that the method of correction employed could have a dramatic and not easily predictable impact on the population dynamics of individual trajectories. \cite{heller_joswig}

Previous work investigating the effect of modifying $\lambda$ on behavior at conical intersections has uncovered similar trends as observed here.\cite{lan_frutos,manthe_koppel} Lan and coworkers, employing multi-level Redfield theory with most of the molecules' internal modes comprising the bath, found $\lambda$ to be a limiting parameter on the rate of internal conversion following vertical excitation of the pyrrole-pyridine complex. Doubling $\lambda$ doubled the rate of internal conversion. \cite{lan_frutos} This limiting behavior of $\lambda$ is consistent with our results which show $\lambda$'s striking influence on quantum yield. Manthe and Koppel investigated the closed system wavepacket dynamics of several different molecules with accessible conical intersections including C$_6$H$_6^+$ and NO$_2$ following vertical excitation. Classes of behavior based on $\lambda$ were assigned with the small $\lambda$ regime showing largely diabatic behavior, the large $\lambda$ regime showing largely adiabatic behavior, and an intermediate regime sometimes resulting in complicated interplays of adiabatic and diabatic effects. \cite{manthe_koppel} These results are more difficult to compare to this work, however a striking change in regime between small and large $\lambda$ is visible in this investigation as well as that of Manthe and Koppel.\cite{manthe_koppel}

\section*{Conclusions}
By generalizing TPT to Lindblad dynamics as a means of characterizing committor probabilities of eigenstates and typical quantum transition pathways hopping through energy space, we have elucidated the effect of diabatic coupling strength on dynamics at conical intersections. In a thermal barrier crossing, a larger $\lambda$ results in a higher rate of population transfer between the metastable and stable well. The committors for each eigenstate of the system and principle paths to analyze for mechanistic information provided by TPT reveal that the energy of the conical intersection itself is never reached during barrier crossing events. Rather, we find that when $\lambda$ is small, deep tunneling is principally responsible for population transfer between the stable and metastable well whereas when $\lambda$ is large, traversing around the conical intersection is principally responsible.
  
By applying a modified TPT approach to treat relaxation following a vertical excitation, we find that increasing $\lambda$ greatly increases the diversity of the relaxation pathways available to the system and results in many quantum jumps whose purpose is merely to dissipate energy prior to commitment to one well or another. Analysis of pathways of principal importance at several $\lambda$ values reveals that the fate of low $\lambda$ trajectories is largely determined during dephasing, meaning that a trajectory bound for eigenstate $L$ typically has committed upon collapse of the wavefunction to an eigenstate, whereas at higher $\lambda$ trajectories typically relax and lose energy until comparable with the adiabatic potential energy barrier before committing. 
  
These fundamental differences in behavior between low and high $\lambda$ systems imply a trade-off. High $\lambda$ systems have a high photoyield, but experience a fast thermal barrier crossing by going around the conical intersection. Low $\lambda$ systems have low photoyields, but must cross the barrier by slow, deep tunneling mechanisms. 
Although limited by the weak coupling assumption, we have nevertheless provided a means of studying reactions in highly nonadiabatic regimes. This work opens avenues of exploration into the dynamics of conical intersections and other systems. This generalization of Transition Path Theory to quantum dynamics shows  promise for elucidating mechanisms in a variety of circumstances, and contributes a useful view of quantum transition pathways.

\section*{Acknowledgements}
We would like to thank Eran Rabani for useful discussions. M.C.A., A.J.S. and D.T.L. were supported by the U.S. Department of Energy, Office of Science, Basic Energy Sciences, CPIMS Program Early Career Research Program under Award DE-FOA0002019.

{\color{black}
 \section*{Data Availability}
The data that support the findings of this study are openly
available in Zenodo at 10.5281/zenodo.6950371\cite{repo}
}
\appendix
\section{Model Details} \label{AppA}
The modified conical intersection based on a pyrazine parameterization \cite{chen_lipeng} is specified in Table \ref{tab1}. {\color{black} The $Q_c$ basis includes 40 unshifted harmonic oscillator basis functions and the $Q_t$ basis includes 110 unshifted harmonic oscillator basis functions save in the case of density matrix propagation initialized in an eigenstate which employs 30 and 80 respectively. Thermal equilibrium calculations include a truncated basis of 700 energy eigenstates save in the case of density matrix propagation initialized in an eigenstate which uses a truncated basis of 240 energy eigenstates.}  Vertical excitation calculations employ a truncated basis of 700 energy eigenstates, {\color{black} save in the case of vertical density matrix propagation which employs a truncated basis of 800 energy eigenstates. A total of 10,000 vertical relaxation} trajectories are simulated at each coupling strength. The temperature for all simulations is 300 K. 

\renewcommand{\arraystretch}{1.5}
\begin{table}
\centering
\begin{tabular}{P{4.7cm} | P{3.4cm} }
\hline 
\hline 
Parameter  & value (eV)   \\
\hline
Tuning mode frequency $\omega_{t} $ & 0.074   \\
Coupling mode frequency $\omega_{c}$ & 0.118  \\
State 1 displacement $\kappa_1 $ & 0.358  \\
State 2 displacement $\kappa_2 $ & -0.315  \\
State 1 energy shift $E_1  $ & 4.21  \\
State 2 energy shift $E_2  $ & 3.94  \\
Reference diabatic coupling $\lambda_0  $ & 0.262  \\
Characteristic bath frequency $\omega_b  $ & 0.01316  \\
Reorganization energy $\eta $ & 2.628 $\times 10^{-4}$  \\\hline 
\hline
\end{tabular}
\caption{Simulation parameters for the linear vibronic coupling model.} 
\label{tab1}
\end{table}

\section{Transition Matrix Details} \label{AppB}
Markov state models for TPT in the thermal case are assembled from $\tau=250 \; \mathrm{au}$ ($6.047 \; \mathrm{fs}$) density matrix calculations initialized in each eigenstate following Eq.~\ref{Eq:Tij}. Max-min flux paths are determined with repeated applications of Dijkstra's algorithm. \cite{dijkstra} A timescale ($\tau$) which is too large risks incorporating large numbers of double jumps into the Markov model thereby skipping over eigenstates in the reactive pathways. For the largest $\lambda$, the principal path length began to decrease, indicating double jumps becoming dominant, at approximately $\tau$=100 fs. For the smallest $\lambda$, a change in dominant path length did not occur until $\tau$=1000 fs. A timescale which is too small leads to numerical problems due to off-diagonal entries in the transition matrix becoming negligible and diagonal entries becoming 1 to machine precision.
\begin{figure}
\begin{center}
\includegraphics[width=8.5cm]{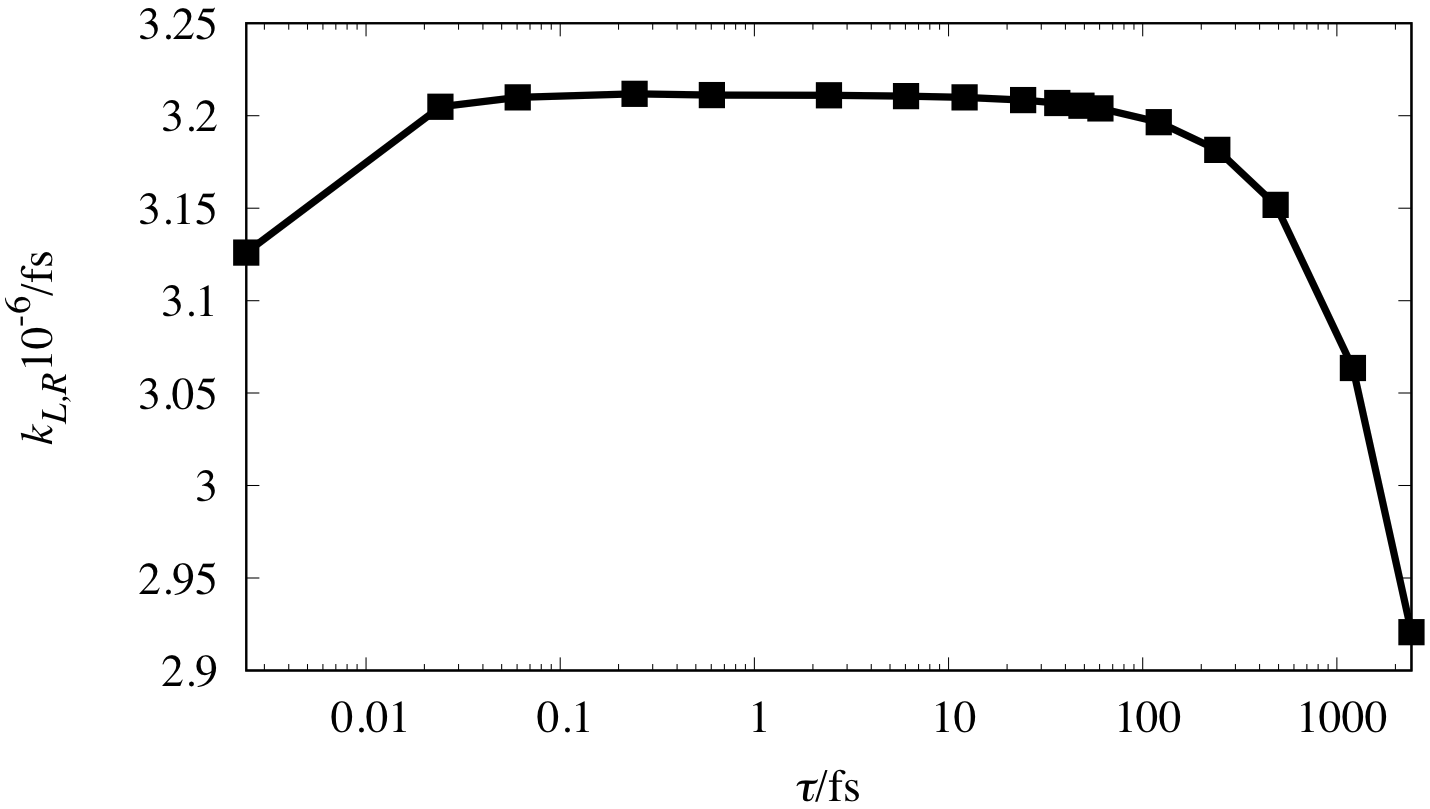}
\caption{Thermal barrier crossing as a function of $\tau$ for the highest coupling, $\lambda=1.5\lambda_0$ case.}
\label{flux_tau_1.5}
\end{center} 
\end{figure}
\begin{figure}
\begin{center}
\includegraphics[width=8.5cm]{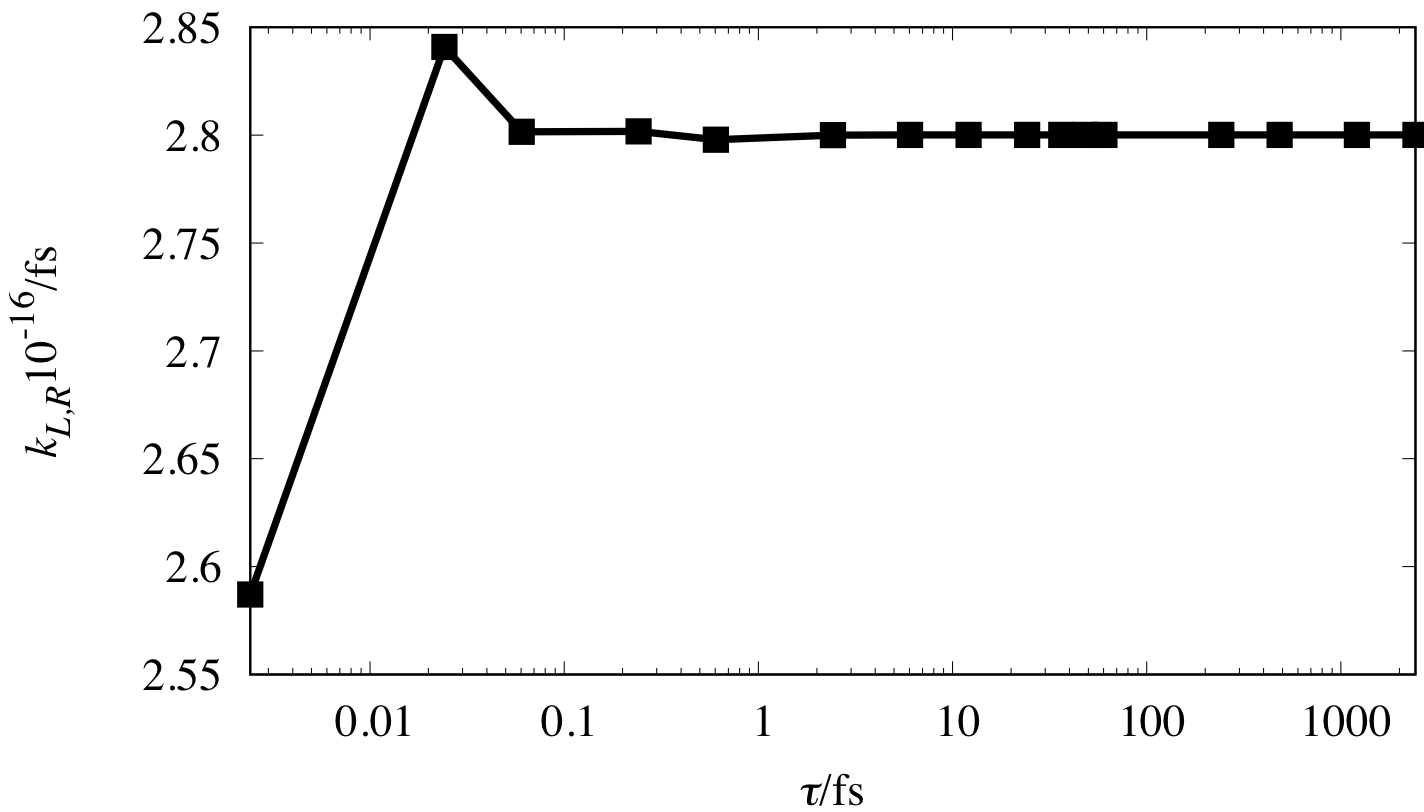}
\caption{Thermal barrier crossing rate as a function of $\tau$ for the lowest coupling, $\lambda=0.05\lambda_0$ case.}
\label{flux_tau_0.05}
\end{center} 
\end{figure}

Figures \ref{flux_tau_1.5} and \ref{flux_tau_0.05} show the thermal barrier crossing rate as a function of $\tau$ for the largest and smallest $\lambda$ in the study respectively. Over many orders of magnitude in $\tau$, the rates remain stable. At very small $\tau$, numerical issues result in instability whereas the largest $\tau$ at high $\lambda$ leads to instability by skipping steps along the reactive pathway. The chosen value of $\tau=6.047 \; \mathrm{fs}$ is unlikely to result in incorporation of double jumps while also avoiding potential numerical problems. The $\tau$ chosen was also verified by recalculating the Markov models at timescales an order of magnitude larger and smaller for the largest and smallest $\lambda$ values and confirming that the thermal reaction rates, committors, and principal pathways were insensitive to the change.

\section*{References}
\bibliography{ref3}
\end{document}